%% file: paper.tex
\RequirePackage{fix-cm}
\documentclass[twocolumn]{svjour3}           
\smartqed  
\usepackage{graphicx}
\usepackage{balance}  
\usepackage{enumitem}

\usepackage{amssymb}
\usepackage{algorithm}
\usepackage{algorithmic}
\usepackage{subfigure}
\usepackage{multirow}
\usepackage{bm}
\usepackage{nicefrac}
\usepackage{subfigure}
\usepackage{caption}
\usepackage{enumitem}
\usepackage{url}
\usepackage{color}
\usepackage{cite}
\usepackage{times}
\usepackage{amsmath}

\usepackage{array}
\newcolumntype{L}[1]{>{\raggedright\let\newline\\\arraybackslash\hspace{0pt}}m{#1}}
\newcolumntype{C}[1]{>{\centering\let\newline  \\\arraybackslash\hspace{0pt}}m{#1}}
\newcolumntype{R}[1]{>{\raggedleft\let\newline \\\arraybackslash\hspace{0pt}}m{#1}}

 \usepackage{mathptmx}      
 
\newcommand{\changed}[1]{\textcolor{black}{#1}}
 
 \newif\ifTKDE

\TKDEfalse

\ifTKDE
\newcommand{\revised}[1]{\textcolor{black}{#1}}
\else
\newcommand{\revised}[1]{\textcolor{black}{#1}}
\fi

\newtheorem{pruning}{Pruning Rule}
\begin{document}

\title{Safest Nearby Neighbor Queries in Road Networks (Full Version)}

\author{Punam Biswas \and Tanzima Hashem \and Muhammad Aamir Cheema}

\institute{Punam Biswas \at
    Department of Computer Science and Engineering,
	Bangladesh University of Engineering and Technology \\
	\email{0417052051@grad.cse.buet.ac.bd}
	\and
	Tanzima Hashem \at
	Department of Computer Science and Engineering,
	Bangladesh University of Engineering and Technology\\
	\email{  tanzimahashem@cse.buet.ac.bd}
	\and
	Muhammad Aamir Cheema\at
    Faculty of Information Technology, Monash University \\
    \email{aamir.cheema@monash.edu} 
}

\date{ }

\maketitle
\begin{abstract}
\revised{Traditional route planning and $k$ nearest neighbors queries  only consider distance or travel time and ignore road safety altogether. However, many travellers prefer to avoid risky or unpleasant road conditions such as roads with high crime rates (e.g., robberies, kidnapping, riots etc.) and bumpy roads. To facilitate safe travel, }we introduce a novel query for road networks called the $k$ safest nearby neighbors ($k$SNN) query.
Given a query location $v_l$, a distance constraint $d_c$ and a point of interest $p_i$,
we define the safest path from $v_l$ to $p_i$ as the path with the highest \emph{path safety score} among all the paths from $v_l$ to $p_i$ with length less than $d_c$. 
The path safety score is computed considering the road safety of each road segment on the path.
Given a query location $v_l$, a distance constraint $d_c$ and a set of POIs $P$, a $k$SNN query returns $k$ POIs with the $k$ highest path safety scores in $P$ along with their respective safest paths from the query location.
We develop two novel indexing structures called $Ct$-tree and a safety score based Voronoi diagram (SNVD). We propose two efficient query processing algorithms each exploiting one of the proposed indexes to effectively refine the search space using the properties of the index. 
Our extensive experimental study on real datasets demonstrates that our solution is on average an order of magnitude faster than the baselines.

\end{abstract}

\input{sections/2_Introduction}

\input{sections/4_1_Related_Works}

\input{sections/3_ProblemFormulation}

\input{sections/4_2_IncrementalNetworkExpansion.tex}

\input{sections/5_CtTreeBasedSolution}

\input{sections/6_VoronoiBasedSolution}

\input{sections/7_Experiments}

\input{sections/8_Conclusion}

\appendix
\section*{Appendix}
\input{sections/9_Appendix}
\bibliographystyle{plain}
\bibliography{references}

\end{document}

%% file: sections/2_Introduction.tex

\section{Introduction}
\label{sec:intro}

\revised{
Crime incidents such as kidnapping and robbery on roads are not unusual especially in developing countries~\cite{stubbert2015crime,natarajan2016crime,spicer2016street}. Street harassment (e.g., eve teasing, sexual assaults) is a common scenario that mostly women experience on roads~\cite{news1, news2}.
A traveler typically prefers to avoid a road segment with high crime or harassment rate. Similarly, during unrest in a country, people prefer to avoid roads with protests or riots.
Also, elderly or sick people may prefer to avoid bumpy roads. 
However, traditional $k$ nearest neighbors ($k$NN) queries that find $k$ closest points of interest (POI) (e.g., a fuel station or a bus stop) fail to consider a traveler's safety or convenience on roads. In real-world scenarios, a user may prefer visiting a nearby POI instead of their nearest one if the slightly longer path to reach the nearby POI is safer than the shortest path to the nearest
POI. In this paper, to allow travelers to avoid different types of inconveniences on roads (e.g., crime incidents, harassment, bumpy roads etc.), we introduce a novel query type, called a \emph{$k$ safest nearby neighbors ($k$SNN) query} and propose novel solutions for efficient query processing.}

\changed{We use safety score of a road segment to denote the associated convenience/safety of traveling on it. In this context, a user may want to avoid roads with low safety scores. Intuitively, a path is safer if it requires a smaller distance to be travelled on the least safe roads. Based on this intuition, we develop a measure of the Path Safety Score (PSS) by considering the safety scores and lengths of the road segments included in a path, and use it to find the safest path between two locations. Since in real-world scenarios, a user may not want to travel on paths that are longer than a user-defined distance constraint $d_c$, we incorporate $d_c$ in the formulation of the $k$SNN query as discussed shortly.} 

\begin{figure}[t]
   \centering
   \begin{tabular}[b]{c}
   \includegraphics[width=1\linewidth]{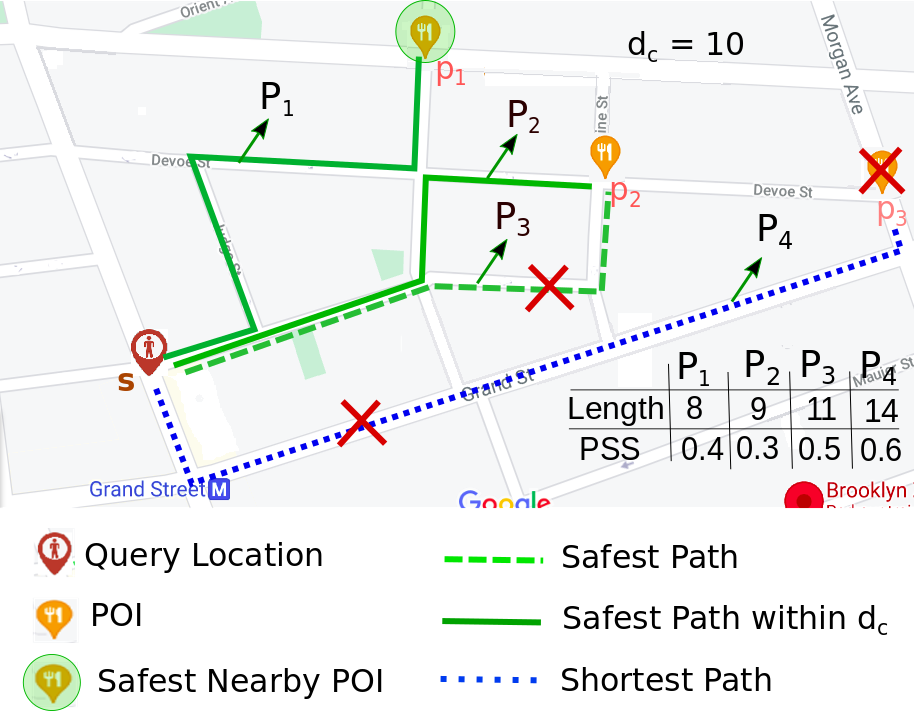} \\
   \end{tabular}
  \caption{An example $k$SNN query for $k=1$ and $d_c=10$.}
   \label{intro:fig}
 \end{figure}

\changed{Given a set of POIs $P$ on a road network, a query location $v_l$ and a distance constraint $d_c$, a $k$SNN query returns $k$ POIs (along with paths from $v_l$ to them) with the highest PSSs considering only the paths with distances less than $d_c$. Consider Fig.~\ref{intro:fig} that shows $3$ restaurants ($p_1$ to $p_3$),  a query located at $s$ and the Path Safety Scores (PSS) of some paths from $s$ to the three POIs (we formally define PSS in Section~\ref{problemformulation}). Assume a $1$SNN query (i.e., $k=1$) located at $s$ and $d_c=10$km. The restaurants that meet the distance constraint are the candidates for the query answer (i.e., $p_1$ and $p_2$).
The POI $p_3$ is not a candidate because the length of the shortest path $P_4$ between $s$ and $p_3$ is 14, which exceeds $d_c$.
The safest path between $s$ and $p_1$ is $P_1$. The PSS of $P_1$ is 0.4 and its length is $8$km, which is smaller than $d_c=10km$. Though the safest path $P_3$ between $s$ and $p_2$ has a higher PSS than that of $P_1$, we do not consider $P_3$ because it is longer than $d_c=10km$. The PSS of the safest path $P_2$ between $s$ and $p_2$ within $d_c$ is 0.3. Thus, $1$SNN query (also called SNN query) returns $p_1$ along with the path $P_1$.}


Finding $k$SNNs in a road network is a computational challenge because the number of POIs in the road network and the number of possible paths between a user's location and a POI can be huge. The processing overhead of a $k$SNN algorithm depends on the amount of required road network traversal and the number of POIs considered for finding the safest nearby POIs. In this paper, we develop two novel indexing structures called \emph{Connected Component Tree ($Ct$-tree)} and \emph{safety score based network Voronoi diagram (SNVD)} to refine the search space and propose two efficient algorithms for finding the safest nearby POIs in road networks. In addition, we exploit incremental network expansion (INE) technique~\cite{PapadiasZMT03} to evaluate $k$SNN queries.

A $Ct$-tree recursively partitions the road network graph into connected and safer subgraph(s) by recursively removing the roads with the smallest safety scores. Each node of the $Ct$-tree represents a subgraph, and for every subgraph, the $Ct$-tree stores some distance related information. By exploiting $Ct$-tree properties and the stored distance information, we develop pruning techniques that avoid exploring unnecessary paths that cannot be the safest ones to reach a POI and have distances less than the distance constraint. Although a number of indexing structures~\cite{Guttman84,BeckmannKSS90,LeeLZT12,ZhongLTZG15,LiCW19} have been developed, they are developed to reduce the processing overhead for finding nearest neighbors and cannot be applied or trivially extended for efficient processing of $k$SNN queries.  

In the past, the Voronoi diagram~\cite{KolahdouzanS04} has also been widely used for finding nearest neighbors with reduced processing overhead. The traditional Voronoi diagram divides the road network graph into subgrpahs such that each subgraph corresponds to a single POI which is guaranteed to be the nearest POI of every query location in this subgraph. We introduce SNVD that guarantees that, for every query location in a subgraph, its \emph{unconstrained safest neighbor} (i.e., SNN where the distance constraint is ignored) is its corresponding POI. Note that the unconstrained safest neighbor of a query location is not necessarily its SNN depending on the distance constraint specified by the user.  By exploiting the SNVD properties, we develop an algorithm to identify the SNNs, and similar to our $Ct$-tree based approach we improve the performance of our SNVD based algorithm with novel pruning techniques to refine the search space. 

Incremental Network Expansion (INE) is among the most efficient nearest POI search techniques that do not rely on any distance-based indexing structure. According to our PSS measure, the PSS of a subpath is greater than or equal to that of the path. This property allows us to adapt the INE based approach for finding $k$SNNs. The INE based approach starts the search from a user's location, progressively expands the search and continues until the required POIs are identified. Our experiments show that our $Ct$-tree and SNVD based algorithms outperform the INE based approach with a large margin. 
 
\revised{The steps of a straightforward solution to evaluate $k$SNNs are as follows: (i) identify candidate POIs which are the POIs with Euclidean distance less than the distance constraint $d_c$ from the query location; (ii) compute the safest paths with distance less than $d_c$ from the query location to each of the candidate POIs; and (iii) select $k$ POIs with the highest path safety scores. This solution would incur very high processing overhead because it requires multiple searches. All of our algorithms find $k$SNNs with a single search in the road network. We consider this straightforward approach as a baseline in our experiments and compare the performance of our algorithms against it.}

To the best of our knowledge, no work has so far addressed the problem of finding $k$SNNs and quantified the PSS measure. Our major contributions are as follows:
\begin{itemize}
    \item We formulate a measure to quantify the PSS by incorporating road safety scores and individual distances associated with the road safety scores. Our solution can be easily extended for any other PSS measure that satisfy certain properties. 
    \item We develop two novel algorithms for processing $k$SNN queries efficiently using a $Ct$-tree and an SNVD, respectively. In addition, we adapt INE based approach for finding the safest nearby POIs.  We propose novel pruning techniques to refine the search space and further improve the performance of our algorithms.
    \item We conduct an extensive experimental study using real datasets that demonstrates that the two proposed algorithms significantly outperform the baseline and the INE based approach.
\end{itemize}

\revised{The rest of this paper is organized as follows. We discuss related work in Section~\ref{relatedwork}. In Section~\ref{problemformulation}, we formally define the $k$SSN query. In Sections~\ref{INE},~\ref{CtTreeBasedSolution} and~\ref{voronoiBasedSolution}, we discuss our solutions based on INE, Ct-tree and SNVD, respectively. Experimental study is presented in  Section~\ref{experiments} followed by conclusions and future work in Section~\ref{conclusion}}.

%% file: sections/4_1_Related_Works.tex
\section{Related Work} \label{relatedwork}

\ifTKDE

\else
\subsection{\revised{Route Planning}}
\fi

\revised{A variety of route planning problems have been studied in the past such as: \emph{shortest route computation}~\cite{Dijkstra59,ZhongLTZG15} which requires to find the path with the smallest overall cost;
\emph{multi-criteria route planning}~\cite{NadiD11,SalgadoCT18} which aims to return routes considering multiple criteria (e.g., length, safety, scenery etc.); 
\emph{obstacle avoiding path planning}~\cite{shen2020euclidean, LambertG03, leenen2012constraint, BergO06, AnwarH17} which returns the shortest path avoiding a set of obstacles in the space; 
\emph{safe path planning}~\cite{AljubayrinQJZHL17, AljubayrinQJZHW15, KimCS14} that aims to return paths that are safe and short;
\emph{alternative route planning}~\cite{li2021comparing,NadiD11,li2020continuously} where the goal is to return a set of routes significantly different from each other so that the user has more options to choose from; \emph{multi-stop route planning} (also called trip planning)~\cite{HashemHAK13,SharifzadehKS08,SalgadoCT18} which returns a route that passes through multiple stops/POIs satisfying certain constraints; and \emph{route alignment}~\cite{SinghSS19} that  requires determining an optimal alignment for a \emph{new} road to be added in the network such that the cost considering different criteria is minimized. Below, we briefly discuss some of these route planning problems most closely related to our work.
}

\noindent
\revised{
\textbf{Multi-criteria route planning.} Existing techniques~\cite{FuLL14, ShahBLC11,NadiD11,SalgadoCT18} typically use a scoring function (e.g., weighted sum) to compute a single score of each edge considering multiple criteria. Once the score of each edge has been computed, the existing shortest path algorithms can be used to compute the path with the best score. In contrast, our work cannot trivially use the existing shortest path algorithms due to the nature of the PSS. 
Also, as noted in~\cite{GalbrunPT16}, it is non-trivial for a user to define an appropriate scoring function combining the multiple criteria. 
Thus, some existing works~\cite{GalbrunPT16,josse2015tourismo} approach the multi-criteria route planning differently and compute a set of \emph{skyline routes}  which guarantees that the routes returned are not \emph{dominated} by any other routes. This is significantly different from our work as it returns possibly a large number of routes for an origin-destination pair instead of the safest route.
}

\noindent
\revised{
\textbf{Obstacle avoiding path planning.} Inspired by applications in robotics, video games, and indoor venues~\cite{cheema2018indoor}  a large number of existing works~\cite{shen2020euclidean, LambertG03, leenen2012constraint, BergO06, AnwarH17,shen2022fast} focus on finding the shortest path that avoids passing through a set of obstacles (e.g., walls) in the space. A recent work aims at finding indoor paths that avoid crowds~\cite{liu2021towards}. Even when obstacles are considered as unsafe regions, these works are different from our work because, in our case, the path may still pass through unsafe edges whereas these works assume that the path cannot cross an obstacle (e.g., a shopper cannot move through the walls). 
}

\noindent
\revised{
\textbf{Safe path planning.}
There are several existing works that consider safety in path planning. In~\cite{AljubayrinQJZHL17, AljubayrinQJZHW15, KimCS14}, the authors divide the space into safe and unsafe zones and minimize the distance travelled through the unsafe zones. However, these works are unable to handle different safety levels of roads and do not consider any distance constraint on the paths. In Section~\ref{problemformulation}, we show that our work is more general and is applicable to a wider range of definitions of safe paths.  Some other existing works~\cite{FuLL14, GalbrunPT16, ShahBLC11} also consider safety, however, they model the problem as multi-criteria route planning and are significantly different from the problem studied in this paper (as discussed above). The most closely related work to our work is~\cite{Islam21} which develops an INE-based algorithm to find the safest paths between a source and a destination by considering individual distances associated with different safety scores of a path. We use this as a baseline in our work. }

\revised{The problem studied in this paper is significantly different from the above-mentioned route planning problems. Unlike route planning problems, our problem does not have a fixed destination, (i.e., each POI is a possible destination) and the goal is to find $k$ POIs with the safest paths. In other words, the problem studied in this work is a POI search problem similar to a $k$ nearest neighbor ($k$NN) query.
In our experimental study, we adapt the state-of-the-art safe path planning algorithm~\cite{Islam21} to compute $k$SNNs and compare against it. Specifically, one approach to find $k$SNNs is to compute the safest path using~\cite{Islam21} for each POI that has Euclidean distance less than $d_c$ from the query location. The application of~\cite{Islam21} to find $k$SNNs requires multiple independent safest path searches and incurs extremely high processing overhead (as shown in our experimental results).
Next, we briefly discuss recent works on $k$NN queries.}

\ifTKDE
\noindent
\revised{
\textbf{$k$NN Queries.} The problem of finding $k$NN queries have been extensively studied, e.g., see~\cite{abeywickrama2016k} which describes and evaluates some of the most notable $k$NN query processing algorithms on road networks. A straightforward application of a existing $k$NN algorithms to $k$SNNs is prohibitively expensive as it would require two independent traversals of the road network for first finding $k^\prime > k$ nearest neighbors as candidate $k$SNNs, and then computing the safest paths from the query location to each of the candidate $k$SNN, respectively. We extend the INE~\cite{PapadiasZMT03} technique to identify $k$SNNs with a single search in the road network.
}
\else

\subsection{$k$ Nearest neighbor Queries}
The problem of finding $k$ nearest neighbors (NNs) has been extensively studied in the literature. Researchers have exploited the incremental network expansion (INE), incremental euclidean restriction (IER), and indexing techniques to solve $k$NN queries. INE~\cite{PapadiasZMT03} progressively explores the road network paths from the query location in order of their minimum distances until $k$ NNs are identified. IER~\cite{PapadiasZMT03} uses the Euclidean lower bound property that the road network distance between two points is always greater than or equal to their Euclidean distance to find $k$ NNs in the road network. Different indexing technique dependent $k$NN algorithms  like Distance Browsing~\cite{SametSA08,SankaranarayananAS05}, ROAD~\cite{LeeLZT12}, G-tree~\cite{ZhongLTZG15} and Voronoi diagram~\cite{KolahdouzanS04} have been developed to refine the search space and reduce the processing overhead. All of the above algorithms except INE are dependent on the distance related properties and cannot be applied or trivially extended for finding $k$ SNNs. An extensive experimental study has been conducted in~\cite{abeywickrama2016k} to compare existing $k$NN algorithms on road networks.

The straightforward application of a NN algorithm for finding $k$ SNNs is prohibitively expensive as it would require two independent traversals of the road network for finding $k^\prime >> k$ nearest neighbors as candidate SNNs, and then computing the safest paths from the query location to each of the candidate SNN, respectively. We extend the INE technique to identify $k$ SNNs with a single search in the road network.

\fi

%% file: sections/3_ProblemFormulation.tex
\section{Problem Formulation} \label{problemformulation}
We model a road network as a weighted graph $G = (V, E)$, where $V$ is a set of vertices and $E$ is a set of edges. A vertex $v_i\in V$ represents a road junction in the road network and an edge $e_{ij}\in E$ represents a road between $v_i$ to $v_j$. Each edge $e_{ij}$ is associated with two \revised{values $w_{ij}^d$ and $w_{ij}^{ss}$}.
\revised{Here, $w_{ij}^d$ is a positive value representing the weight of the edge $e_{ij}$, e.g., length, travel time, fuel cost etc. For simplicity, hereafter, we use length to refer to  $w_{ij}^d$.
$w_{ij}^{ss}$ represents the edge safety score (\revised{ESS}) of $e_{ij}$ (higher the safer). Computing $w^{ss}_{ij}$ is beyond the scope of this paper and we assume safety scores are given as input (e.g., computed using existing geospatial crime mapping approaches~\cite{chainey2013gis}).  Table~\ref{table:two} summarizes the main notations used the paper.}

A path $pt_{ij}$ between two vertices $v_i$ and $v_j$ is a sequence of vertices such that the path starts at $v_i$, ends at $v_j$, and an edge exists between every two consecutive vertices in the path. \revised{Cost of a path $pt_{ij}$ is the sum of the weights of the edges in the path, e.g., total length, total travel time etc. For simplicity, hereafter, we use distance/length to refer to the cost of a path $pt_{ij}$ and denote it as  $dist(pt_{ij})$.
Let $d_c$ be a user-defined distance constraint. We say that a path $pt_{ij}$ is valid if $dist(pt_{ij}) < d_c$.} 

    \begin{table}[t]
    \centering
    \caption{A list of notations} \label{table:two}
    
    \begin{tabular}{|p{1.5cm}|p{5.5cm}|}
    \hline
        \textbf{Notation} &\textbf{Explanation}  \\ \hline
           $G(V,E)$ & The road network graph  \\ \hline
           $e_{ij}$ & An edge connecting vertices $v_i$ and $v_j$\\ \hline
           $w_{ij}^d$ & \revised{Weight (e.g., length, travel time) of $e_{ij}$}  \\ \hline
           $w_{ij}^{ss}$ & \revised{Edge safety score (ESS) of $e_{ij}$}   \\ \hline
           $s^{max}$ & \revised{Maximum ESS, i.e., $argmax_{e_{ij}\in E}~w_{ij}^{ss}$}   \\ \hline
           $v_l$ & A query location \\ \hline
           $d_c$ & \revised{A user-defined distance constraint} \\ \hline
           $p_i$ & A POI located at $v_i$ \\ \hline
           $pt_{ij}$ & A path from a vertex $v_i$ to a vertex $v_j$ \\ \hline
           $dist(pt)$ & \revised{Sum of weights of edges in $pt$} \\ \hline
           $pt_{ij}.d_s$ & \revised{Total weight of edges in $pt_{ij}$ with ESS equal to $s$} \\ \hline
           $pss(pt)$ & \revised{Path safety score of $pt$} \\ \hline
           $pt^{sf}_{ij}$ & The safest path between $v_i$ and $v_j$ that has distance less than $d_c$\\ \hline
           $pt^{sh}_{ij}$ & The shortest path between $v_i$ and $v_j$ \\ \hline
           
               \end{tabular}
  
    \end{table}

\subsection{Path Safety Score (PSS)}

\changed{ 
Intuitively, we want to define a Path Safety Score (PSS) measure that ensures that a path that requires smaller distance to be travelled on less safe roads has a higher PSS. Before formally representing this requirement using Property~\ref{ss_property0_1}, we first define $s$-distance (denoted as $d_s$) of a path which is the total distance a path requires  travelling on edges with safety score equal to $s$.}

\begin{figure}[t]
  \centering
  \begin{tabular}[b]{c}
  \includegraphics[width=1\linewidth]{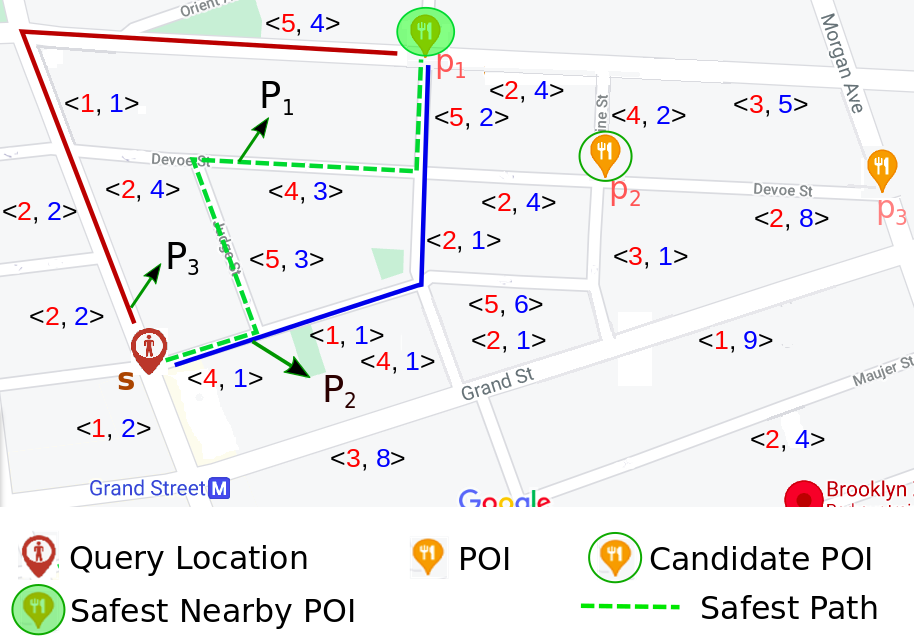} \\
  \end{tabular}
 
  \caption{\revised{For each road segment, its ESS and length (in km) are shown as $\langle${ESS}, {length}$\rangle$. Assuming $d_c=10$km, $p_1$ is the $1$SNN for query $s$ and $P_1$ is the safest path from $s$ to $p_1$.}}
  \label{fig:querySampleOutput}
\end{figure}

\begin{definition} \label{def_ds}
\revised{ $s$-distance: Given a path $pt_{ij}$ and a positive integer $s$, $s$-distance of the path (denoted as $pt_{ij}.d_s$) is the total length of the edges in $pt_{ij}$ which have safety score equal to $s$, i.e., $pt_{ij}.d_s = \sum_{e_{xy}\in pt_{ij} \wedge w_{xy}^{ss}=s} w^d_{xy}$.}
\end{definition}
 \revised{Consider the example of Fig.~\ref{fig:querySampleOutput} that shows three paths $P_1$, $P_2$ and $P_3$ from source $s$ to a POI $p_1$, with lengths $9$, $5$ and $9$, respectively. In Fig.~\ref{fig:querySampleOutput}, $4$-distance of $P_1$ is $P_1.d_4=4$ because there are two edges on $P_1$ with ESS equal to $4$ and their total length is $1+3=4$. Similarly, $P_1.d_1=0$, $P_2.d_1=1$, $P_3.d_1=1$, $P_2.d_2=1$, and $P_3.d_2=2+2=4$.}


\begin{property}
\label{ss_property0_1}
 \revised{Let $pt_{ij}$ and $pt^{\prime}_{ij}$ be two valid paths (i.e., have distances less than $d_c$). Let $s$ be the smallest positive integer for which $pt_{ij}.d_s\neq pt^{\prime}_{ij}.d_s$. The path $pt_{ij}$ must have a higher PSS than $pt^{\prime}_{ij}$ if and only if $pt_{ij}.d_s < pt^{\prime}_{ij}.d_s$.}
\end{property}

\revised{In Fig.~\ref{fig:querySampleOutput}, according to Property~\ref{ss_property0_1}, $P_1$ must have a higher PSS than $P_2$ and $P_3$ because $P_1.d_1=0$ is smaller than $P_2.d_1=P_3.d_1=1$, i.e., $P_1$ does not require traveling on an edge with safety score $1$ whereas $P_2$ and $P_3$ require traveling $1$km on roads with safety score $1$. $P_2$ must have a higher PSS than $P_3$ because, although $P_2.d_1=P_3.d_1$, $P_2.d_2=1$ is smaller than $P_3.d_2=4$. Hereafter, we use $d_s$ to denote $pt_{ij}.d_s$ wherever the path $pt_{ij}$ is clear by context. 
}

\revised{Next, we define a measure of PSS (Definition~\ref{def_PSS}) which satisfies Property~\ref{ss_property0_1}. For this definition, we assume that, for each edge,  $w^d_{ij}\geq 1$ and $w^{ss}_{ij}$ is a positive integer. These assumptions do not limit the applications because $w^d_{ij}\geq 1$ can be achieved by appropriately scaling up if needed and, in most real-world scenarios, safety scores are integer values based on safety ratings (e.g., 1-10).
}

\begin{definition} \label{def_PSS} {Path Safety Score (PSS):} \label{def1} \revised{
Let $d_c>1$ be the distance constraint and $pt_{ij}$ be a valid path (i.e., $dist(pt_{ij})<d_c$). 
Let ${s}^{max}$ be the maximum \revised{ESS} of any edge in the road network $G$.
The PSS $pss(pt_{ij})$ of $pt_{ij}$ is computed as \begin{math}\frac{1}{\sum_{s=1}^{s^{max}}{w_s \times d_s}} \end{math}, where $w_s = d_c^{(s^{max}-s)}$. }
\end{definition}

Here $w_s$ represents the weight (importance) assigned to the edges with \revised{ESS} equal to $s$. 
For example, if $d_c=10$ and $s^{max}=5$, we have $w_1=10^4$, $w_2=10^3$, $w_3=10^2$, $w_4=10^1$ and $w_5=10^0$. \revised{We calculate PSS for paths $P_1, P_2$ and $P_3$ of Figure \ref{fig:querySampleOutput} as follows. 
$d_s$ values for $P_1$ are $P_1.d_1=P_1.d_2=P_1.d_3=0$, $P_1.d_4=4$ and $P_1.d_5=5$. The PSS for path $P_1$ is \begin{math}\frac{1}{(w_4 \times 4) +(w_5 \times 5)}=\frac{1}{(10 \times 4)+(1 \times 5)}= \frac{1}{45}\end{math}. 
For $P_2$, $d_s$ values are $P_2.d_1=1$, $P_2.d_2=1$, $P_2.d_3=0$, $P_2.d_4=1$, $P_2.d_5=2$. The PSS for path $P_2$ is \begin{math}\frac{1}{(w_1 \times 1) +(w_2 \times 1) + (w_4 \times 1) + (w_5 \times 2 )}=\frac{1}{(10^4 \times 1)+(10^3 \times 1)+(10 \times 1)+(1 \times 2)}=  \frac{1}{11012}\end{math}. Finally, $d_s$ values for $P_3$ are $P_3.d_1=1$, $P_3.d_2=4$, $P_3.d_3=0$, $P_3.d_4=0$, $P_3.d_5=4$. The PSS for  $P_3$ is \begin{math}\frac{1}{(w_1 \times 1) + (w_2 \times 4)  +(w_5 \times 4) }=\frac{1}{(10^4 \times 1) + (10^3 \times 4)+(1 \times 4)}= \frac{1}{14004}\end{math}. Thus, as required by Property~\ref{ss_property0_1}, $P_1$ has the highest PSS followed by $P_2$ and then $P_3$.}

Note that the importance $w_s$ is set such that it fulfils the following condition: a unit length edge with \revised{ESS} equal to $s$ contributes more in the sum than all other edges \revised{in a valid path} with \revised{ESS} greater than $s$, i.e., $w_s > dist(pt_{ij})\times w_{s+x}$ for $x\geq 1$ (this is because $dist(pt_{ij})<d_c$ and $w_{s+x}=w_s/d_c^x$). This condition ensures that our PSS measure satisfies Property~\ref{ss_property0_1}. \revised{For example, when $d_c=10$, we have $w_1=10^4$ and $w_2=10^3$. Since each edge has length at least $1$, this implies that $w_1\times d_1 > w_2\times dist(pt_{ij})$ because length of $pt_{ij}$ is less than $d_c=10$.}

Although the PSS of a path can vary depending on $d_c$, the relative ranking of two paths based on PSSs remains the same irrespective of the value of $d_c$ as long as both paths are valid (i.e., have distance less than $d_c)$. 

\subsection{$k$SNN Query}

\revised{We denote the safest valid path between two vertices $v_i$ and $v_j$ as $pt^{sf}_{ij}$ which is a path with the highest PSS among all valid paths from $v_i$ to $v_j$. Now, we define $k$SNN query.}

\begin{definition}\emph{\revised{{A $k$ Safest Nearby neighbors ($k$SNN) Query:} Given a weighted road network $G(V, E)$, a set of POIs $P$, a query location $v_l$ and a distance constraint $d_c$,  a $k$SNN query returns a set $P_k$ containing $k$ POIs along with the safest valid paths from $v_l$ to each of these POIs such that,
for every $p_i\in P_k$ and every ${p_i}^\prime\in P\setminus P_k$,  $pss({pt^{sf}_{li}}) \geq pss(pt^{sf}_{li^\prime})$ where $pss({pt^{sf}_{li}})$ and $pss(pt^{sf}_{li^\prime})$ represent the safest valid paths from $v_l$ to $p_i$ and ${p_i}^\prime$, respectively.}}
\end{definition} 

\changed{In Fig.~\ref{fig:querySampleOutput}, $P_1$ has the highest PSS among all valid paths from $s$ to any of the three POIs,  Thus a $1$SNN query returns $p_1$ along with the path $P_1$ as the answer. }\revised{In some cases, there may not be $k$ POIs whose distances from the query location are less than $d_c$. In such scenario, a $k$SNN query returns all POIs with distances less than $d_c$, ranked according to their PSSs. Hereafter, a $1$SNN query (i.e., $k=1$) is also simply referred as a SNN query.} Following most of the existing works on POI search, we assume that the POIs and query location lie on the vertices in the graph.

\subsection{Generalizing the Problem}

\changed{
We remark that our definition of PSS is  more general than the definitions in previous works~\cite{AljubayrinQJZHL17, AljubayrinQJZHW15, KimCS14} that do not have any distance constraint and  treat each road segment either as safe or unsafe instead of assigning different safety scores to each road as in our work. Consequently, our definition is more general and we can easily use our definition to compute the path that minimizes the distance travelled through unsafe zones (as in~\cite{AljubayrinQJZHL17, AljubayrinQJZHW15, KimCS14}) by assigning each edge in the safe zone a safety score $x$ and each edge in the unsafe zone a safety score $y$ such that $x>y$ and setting $d_c$ to be larger than the length of the longest path in the network.}

\changed{Although our definition of PSS is already more general than the previous work, in this section, we show that our algorithms can be immediately applied to a variety of other definitions of PSS. This further generalizes the problem studied in this paper and the proposed solutions.
Specifically, we propose two algorithms to solve $k$SNN queries and both algorithms are generic in the sense that they do not only work when PSS is defined using Definition~\ref{def_PSS} but are also immediately applicable to a variety of other definitions of PSS. Specifically, our $Ct$-tree based algorithm works for any other definition of PSS as long as it satisfies both of the Properties~\ref{ss_property1} and~\ref{ss_property1_1}, whereas, our SNVD based algorithm is immediately applicable to any definition of PSS as long as it satisfies Property~\ref{ss_property1}.}

\begin{property}
\label{ss_property1}
  \revised{Let $pt_{xy}$ be a subpath of a valid path $pt_{xz}$. The PSS computed using the defined measure must satisfy  $pss(pt_{xz}) \leq pss(pt_{xy})$.}
\end{property}

\changed{Property~\ref{ss_property1} requires that a subpath must have equal or higher PSS than any path that contains this subpath. This requirement is realistic as the risk/inconvenience associated with a path inherits the risk/inconvenience associated with travelling on any subpath of this path.}

\begin{property}
\label{ss_property1_1}
\revised{Let $pt_{xy}.minSS$ be the minimum safety score among all edges in the path $pt_{xy}$, i.e., $pt_{xy}.minSS= argmin_{e_{ij}\in pt_{xy}} w^{ss}_{ij}$. Given two valid paths $pt_{xy}$ and $pt^{\prime}_{xy}$ such that $pt_{xy}.minSS<pt^{\prime}_{xy}.minSS$, the PSS computed using the defined measure must satisfy $pss(pt_{xy}) < pss( pt^{\prime}_{xy})$.}
\end{property}

\changed{Property~\ref{ss_property1_1} is also realistic as it requires that if the least safe edge on a path $P$ is safer than the least safe edge on another path $P'$ then the PSS of $P$ must be no smaller than that of $P'$.}

\revised{Many intuitive definitions of PSS satisfy the above-mentioned properties. For instance, assume that ESS represents the probability that no crime will occur on this edge (as in~\cite{GalbrunPT16}). If PSS is defined as multiplication of \revised{ESS} on all edges on a path~\cite{GalbrunPT16} (i.e., probability that no crime will occur on the whole path), this definition of PSS satisfies Property~\ref{ss_property1} but does not satisfy Property~\ref{ss_property1_1}. Thus, SNVD based algorithm can be used to handle queries involving such PSS. On the other hand, if PSS corresponds to the minimum \revised{ESS} on any edge on the path, then this definition of PSS satisfies both properties, therefore, both Ct-tree and SNVD based approaches can be used.} 

\changed{We remark that our PSS measure (Definition~\ref{def_PSS}) satisfies both properties and thus, both of our algorithms can be used for it. For example, in Fig.~\ref{fig:querySampleOutput}, the PSS for path $P_1$ is \begin{math}\frac{1}{(w_4 \times 4) +(w_5 \times 5)}=\frac{1}{(10 \times 4)+(1 \times 5)}= \frac{1}{45}\end{math}. Now if we consider the subpath of $P_1$ by removing the last edge with ESS 5 and length 2, the PSS increases to \begin{math}\frac{1}{(w_4 \times 4) +(w_5 \times 3)}=\frac{1}{(10 \times 4)+(1 \times 3)}= \frac{1}{43}\end{math} (which satisfies Property~\ref{ss_property1}). Similarly, $P_1.minSS=4$ and $P_2.minSS=1$, and the PSSs of $P_1$ and $P_2$ are $\frac{1}{45}$ and $\frac{1}{11012}$, respectively (which satisfies Property~\ref{ss_property1_1}).} 

\revised{For the ease of presentation, in our problem setting, we assume that each edge has a single safety score $w_{ij}^{ss}$. However, if an edge has multiple safety scores representing the road safety conditions at different times of the day (e.g., day vs night) or different types of crimes, our solutions can be immediately applied by only considering the relative safety score for each edge (e.g., safety scores for night time if the query is issued at night).}

%% file: sections/4_2_IncrementalNetworkExpansion.tex
\section{Incremental Network Expansion}{\label{INE}}

Since the PSS of a path is no larger than that of its subpath (Property~\ref{ss_property1}), we can  adapt the incremental network expansion (INE) search~\cite{PapadiasZMT03} to find $k$SNNs.
Starting from the query location $v_l$, the INE based search explores the adjacent edges of $v_l$. For each edge $e_{li}$, a path consisting of $v_l$ and $v_i$ is enqueued into a priority queue $Q_p$. The entries in $Q_p$ are ordered in the descending order based on their PSSs. Then the search continues by dequeueing a path from $Q_p$ and repeating the process by exploring adjacent edges of the last vertex of the dequeued path. \revised{Before enqueueing a new valid path $p_{lj}$ into $Q_p$, its PSS is incrementally computed from the PSS of the dequeued path $p_{li}$ and the PSS of the path $p_{ij}$ that consists of a single edge $e_{ij}$ \changed{using the following lemma. The proof of Lemma~\ref{ss_property2} is shown in Appendix~\ref{appendix1}.}}

\begin{lemma}
\label{ss_property2}
    \revised{Let $pt_{lj}$ be a valid path such that $pt_{lj}=pt_{li}\oplus pt_{ij}$ where $\oplus$ is a concatenation operation. Then, 
    $pss(pt_{lj}) = \frac{1}{\frac{1}{pss(pt_{li})} + \frac{1}{pss(pt_{ij})}}$.}
\end{lemma}

 
The first SNN is identified once the last vertex of a dequeued path from $Q_p$ is a POI and the distance of the path is less than $d_c$. The search for $k$SNNs terminates when paths to $k$ distinct POIs have been dequeued from $Q_p$ and the distances of the paths are smaller than $d_c$. 

To refine the search space, we check if a path can be pruned before enqueueing it into $Q_p$ using the following pruning rules. Pruning Rule~\ref{lemma:pruning1} is straightforward as it simply uses the distance constraint for the pruning condition.

\begin{pruning}
\label{lemma:pruning1}
    A path $pt_{lj}$ can be pruned if $dist(pt_{li}) \geq d_c$, where $d_c$ represents the distance constraint.
\end{pruning}

\changed{A path $pt_{lj}$ between $v_l$ and $v_j$ can be pruned if there is already a dequeued path $pt^\prime_{lj}$  which is at least as short as $pt_{lj}$ \emph{and} at least as safe as $pt_{lj}$. The following lemma justifies the above intuition. The proof of Lemma~\ref{lemma:pruning} is shown in Appendix~\ref{appendix2}.}
\changed{
\begin{lemma}
\label{lemma:pruning}
If path $pt^\prime_{lj}$ is at least as short as path $pt_{lj}$ and at least as safe as $pt_{lj}$, then path $pt^\prime_{lk}=pt^\prime_{lj}\oplus pt_{jk}$ is at least as short as and at least as safe as path $pt_{lk}=pt_{lj}\oplus pt_{jk}$.
\end{lemma}
}


\changed{Since the INE based search dequeues paths in descending order of PSSs from $Q_p$ and Pruning Rule~\ref{lemma:pruning2} is applied to a path $pt_{lj}$ before enqueueing it into $Q_p$, the PSS of any dequeued path is higher than $pss(pt_{lj})$. Thus, Pruning Rule~\ref{lemma:pruning2} only checks whether there is a dequeued path to $v_j$ that is at least as short as $pt_{lj}$.}


\begin{pruning}
\label{lemma:pruning2}
    A path $pt_{lj}$ can be pruned if $dist(pt_{lj}) \geq D_{sh}[j]$, where $D_{sh}[j]$ represents the distance of the shortest path $pt^\prime_{lj}$ from $v_l$ to $v_j$ dequeued so far.
\end{pruning}

\revised{We remark that, since $D_{sh}[j]$ is less than $d_c$ when a valid dequeued path to $v_j$ exists, Pruning Rule~\ref{lemma:pruning2} facilitates the pruning of the valid paths (i.e., the path length is smaller than $d_c$). On the other hand, Pruning Rule~\ref{lemma:pruning1} prunes the invalid paths, when there is no existing dequeued path that ends at $v_j$, i.e., $D_{sh}[j]=\infty$.}
\ifTKDE
\revised{
We provide complexity analysis of the algorithm in the full version~\cite{biswas2021safest}.}

\else

\revised{\emph{Complexity Analysis.} Let $n_p$ be the total number of valid paths (i.e., with length less than $d_c$) from $v_l$ to any node in the road network graph $G$. Then, the worst case time complexity for finding $k$ SNNs by applying the INE based search using a priority queue is $O(n_p \log(n_p))$. The number of total paths reduces when we apply Pruning Rule~\ref{lemma:pruning2}. Let $r$ be the effect factor of the  Pruning Rule~\ref{lemma:pruning2}, i.e., the pruning rule reduces the number of paths by a factor $r$. Thus, by applying the pruning rule the worst case time complexity of the INE based search becomes $O(\frac{n_p}{r} \log(\frac{n_p}{r}))$. The checking of whether a path can be pruned takes constant time, and it can be  shown that the number of paths for which the pruning rules are applied is $O(\frac{n_p}{r})$ (because INE incrementally explores the search space).}

\fi

%% file: sections/5_CtTreeBasedSolution.tex
\section{Ct-tree}\label{CtTreeBasedSolution}
In this section, we introduce a novel indexing structure, \textbf{\underline{C}}onnected componen\textbf{\underline{t}} tree ($Ct$-tree) and develop an efficient solution based on $Ct$-tree to process $k$SNN queries. Index structures like $R$-tree~\cite{Guttman84}, Contraction Hierarchy (CH)~\cite{GeisbergerSSD08}, Quad tree~\cite{FinkelB74}, ROAD~\cite{LeeLZT12} and $G$-tree~\cite{ZhongLTZG15} have been proposed for efficient search of the query answer based on the distance metric and are not applicable for $k$SNN queries. Some of these indexing techniques~\cite{Guttman84,FinkelB74} divide the space into smaller regions based on the position of the POIs, whereas some other indexing techniques~\cite{GeisbergerSSD08,LeeLZT12} divide the space based on the properties of the road network graph. These existing indexing techniques cannot be applied or trivially extended to compute $k$SNNs because they do not incorporate \revised{ESS}s of the edges in the road network graph.

\begin{figure*}[t!]
   \begin{center}
        \begin{tabular}{cc}
            \hspace{-6mm}
             \includegraphics[width=0.5\textwidth]{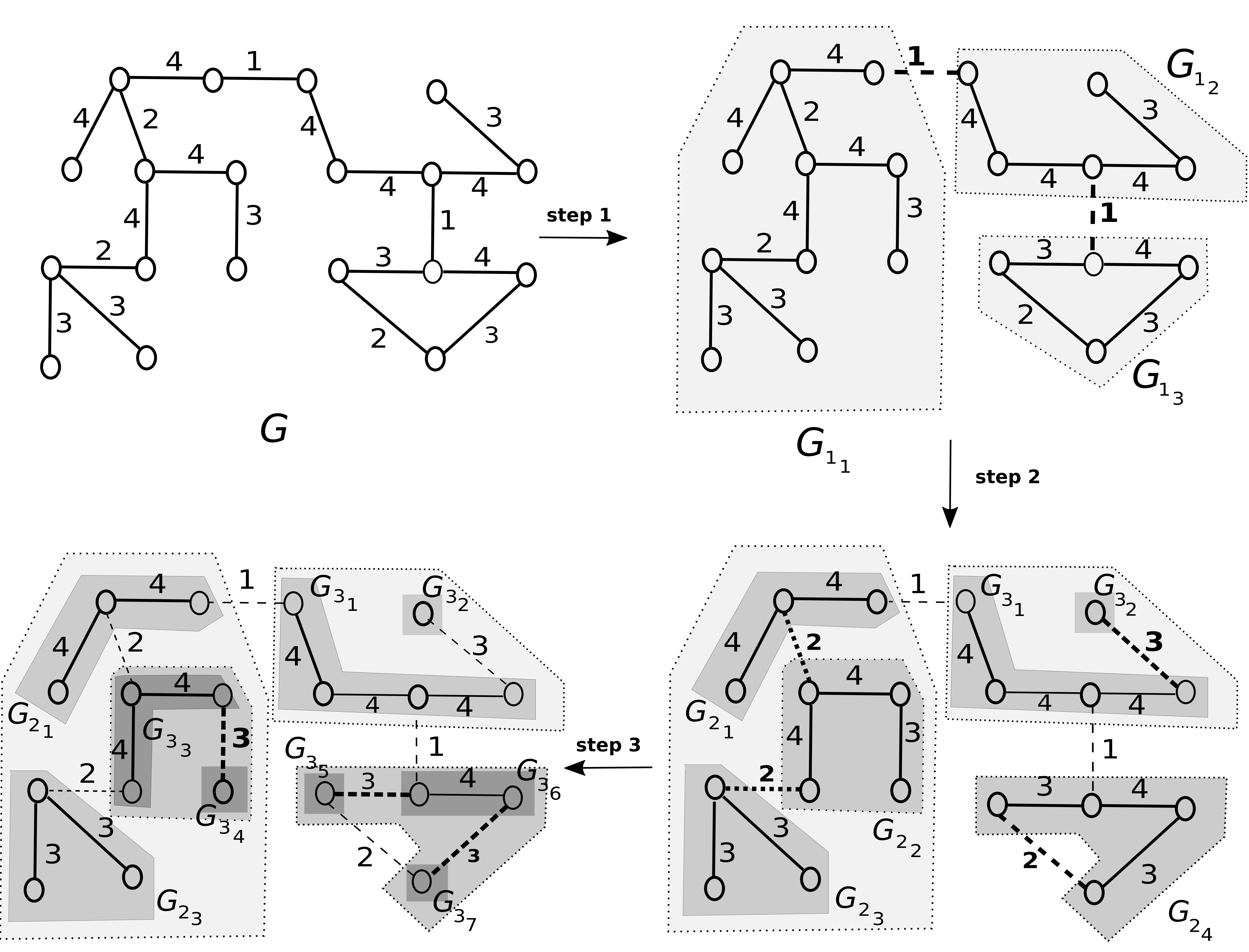}&
             \includegraphics[width=0.5\textwidth]{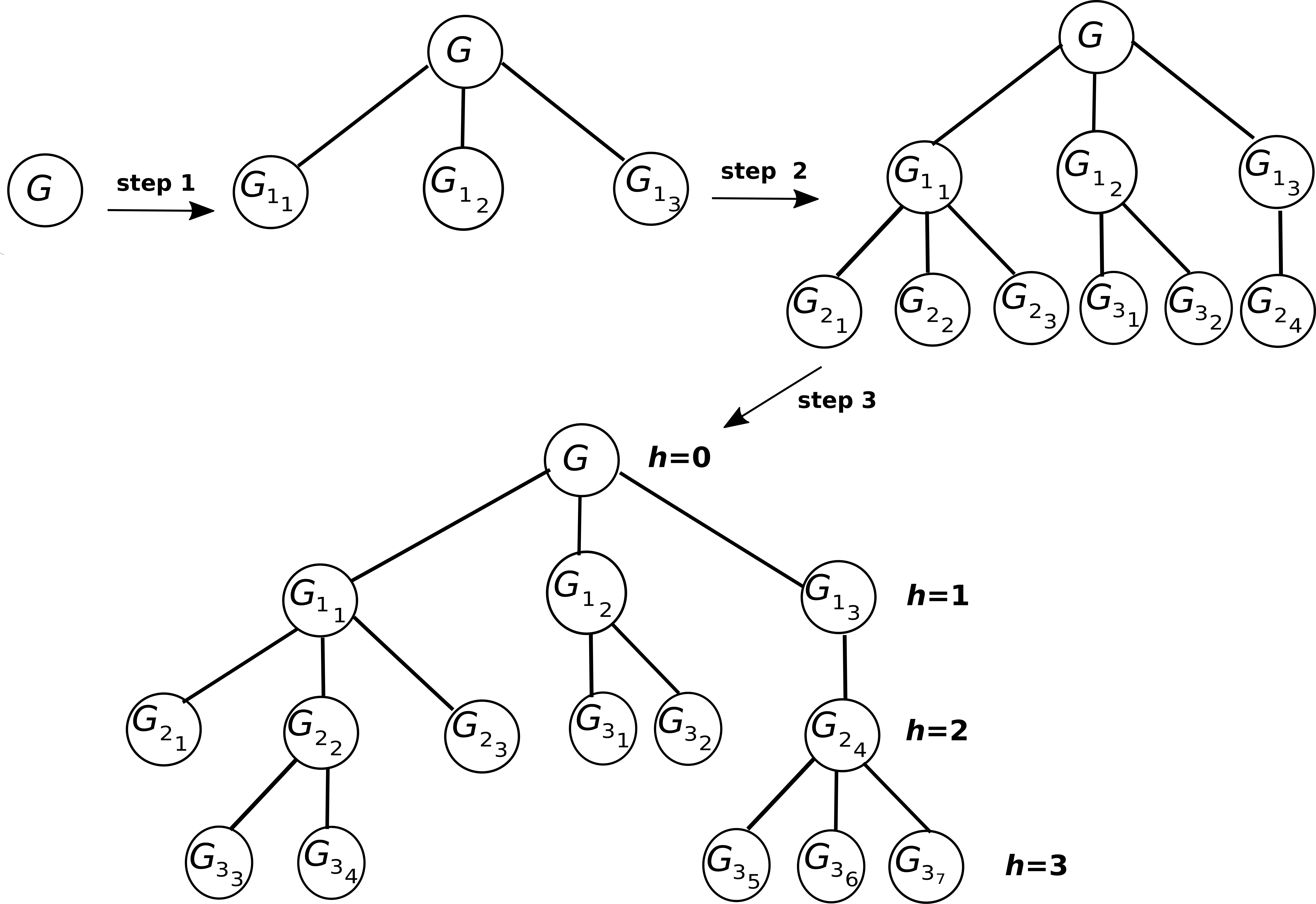}\\ 
             \hspace{-6mm}
            \hspace{10mm} \scriptsize{(a) Ct-tree subgraphs \textsc{}\hspace{10mm}}  &  \scriptsize{(b) Ct-tree \textsc{}\hspace{0mm}}\\
        \end{tabular}
        \caption{Ct-tree Construction Steps}
      \label{fig:Ct_steps} 
    \end{center}
\end{figure*}

\subsection{Ct-Tree Construction and Properties}{\label{construction}}
The key idea to construct a $Ct$-tree is to recursively partition the graph by removing the edges with the smallest \revised{ESSs} in each step. 
Removing the edges with the smallest \revised{ESS} $s$ may partition the graph into one or more connected components. A component is denoted as $G_{s_t}$, where $t$ is a unique identifier for the partitions created by removing edges with \revised{ESS} $s$. Each connected component $G_{s_t}$ is recursively partitioned by removing the edges with the smallest \revised{ESS} within $G_{s_t}$. The recursive partitioning stops when $G_{s_t}$ either contains a single vertex or all edges within $G_{s_t}$ have the same \revised{ESS}.

Without loss of generality, we explain the $Ct$-tree construction process using an example shown in Figure~\ref{fig:Ct_steps}. The original graph has edge \revised{ESS}s in the range of 1 to 4, and for the sake of simplicity, we do not show the edge distances in the figure. The root of the $Ct$-tree represents the original graph $G$. After removing the smallest edge \revised{ESS} 1, $G$ is divided into three connected components $G_{1_1}$, $G_{1_2}$ and $G_{1_3}$. These connected components are represented by three child nodes of the root node at tree height $h=1$. Each of these components is then recursively partitioned by removing the edges with the smallest \revised{ESS}. For example, the edges with \revised{ESS} 2 are removed from $G_{1_1}$, and $G_{1_1}$ is divided into three connected components $G_{2_1}$, $G_{2_2}$ and $G_{2_3}$. The recursive partitioning of $G_{2_1}$ and $G_{2_3}$ stop as they have edges with same \revised{ESS}. On the other hand, after removing the edges with \revised{ESS} 3, $G_{2_2}$ is divided into  $G_{3_3}$ and $G_{3_4}$.

We formally define a $Ct$-tree as follows:
\begin{definition}{\textbf{$Ct$-tree:}}
A $Ct$-tree $C$ is a connected component based search tree, a hierarchical structure that has the following properties:
    \begin{itemize}
    \item The $Ct$-tree root node represents the original graph $G$.
    \item Each internal or leaf $Ct$-tree node represents a connected component $G_{s_t}$, where $G_{s_t}$ does not include any edge with \revised{ESS} smaller than or equal to $s$ and $G_{s_t}$ is included in the graph represented by its parent node.
    \item The maximum height of the tree, ${h}^{max}={s}^{max}-1$ 
    \item Each internal or leaf $Ct$-tree node maintains the following information: the number of POIs $n(G_{s_t})$ in $G_{s_t}$, a border vertex set $B_{s_t}$, the minimum border distance $d_B^{min}(v_x, G_{s_t})$ and the minimum POI distance $d_p^{min}(v_x, G_{s_t})$ for each border vertex $v_x \in B_{s_t}$.s
    \end{itemize}
 \end{definition}

\begin{figure}[htbp]
    \centering
    \includegraphics[width=.45\textwidth]{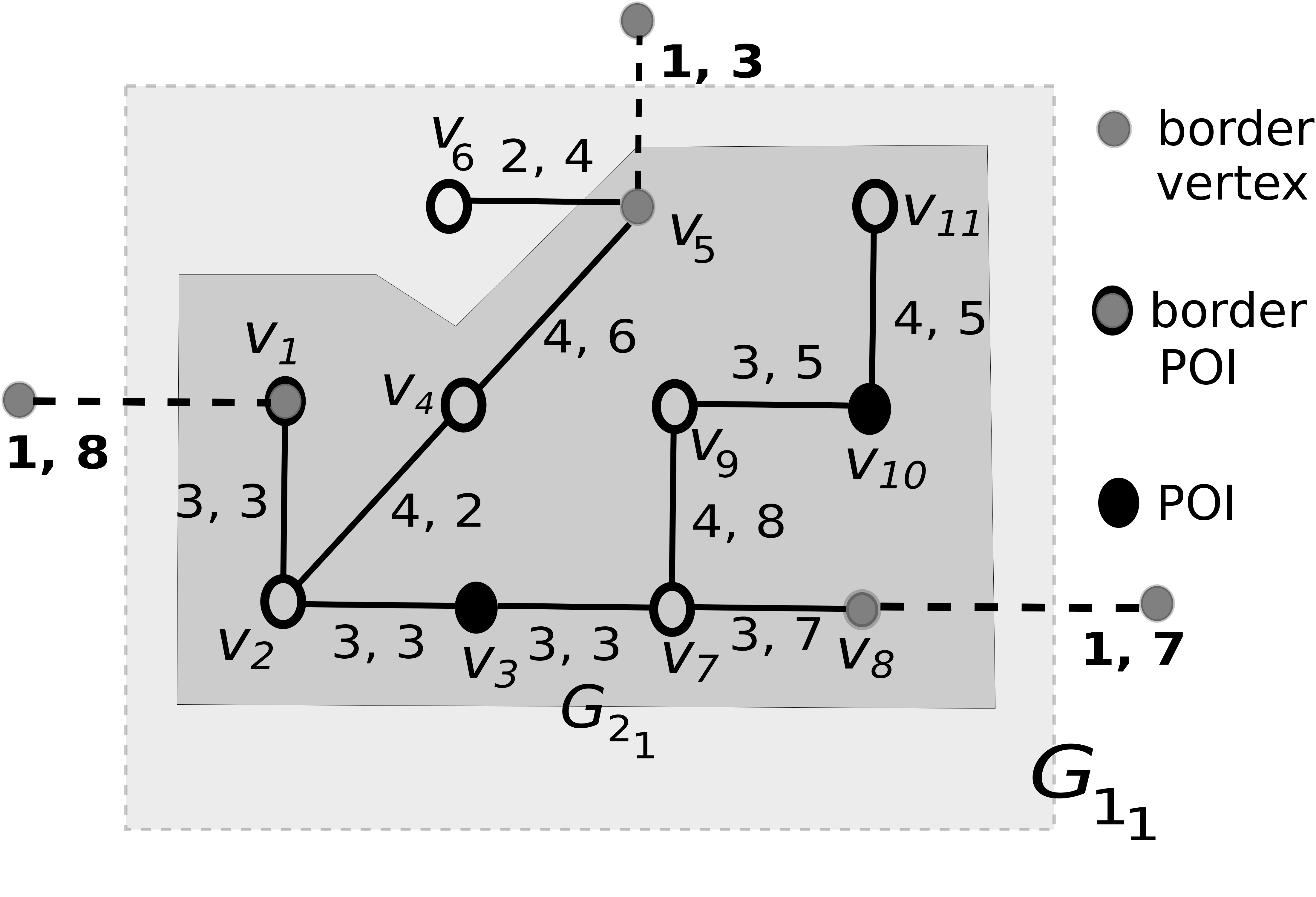}
    \caption{Border vertices, the minimum border distance and the minimum POI distance}
    \label{treeNode}
\end{figure}

\textbf{\emph{Border vertices, the minimum border distance and the minimum POI distance.}} 
A vertex $v_x$ is called a border vertex of a subgraph $G_{s_t}$, if there is an outgoing edge from $v_x$ whose \revised{ESS} is smaller than or equal to $s$ and the edge is not included in $G_{s_t}$. We denote the set of border vertices of $G_{s_t}$ with $B_{s_t}$. For example, in Figure~\ref{treeNode}, $B_{1_2}$ includes $\{v_1, v_5, v_8\}$. A border vertex of $G_{s_t}$ represented by a $Ct$-tree node is also a border vertex of the subgraphs represented by its descendent nodes. For example, $v_5$ is a border vertex of both $G_{1_2}$ and $G_{2_1}$.

For each border vertex, the corresponding $Ct$-tree node stores the minimum border distance and the minimum POI distance. The minimum border distance $d_B^{min}(v_x, G_{s_t})$ of a border vertex $v_x$ of $G_{s_t}$ is defined as the minimum of the distances of the shortest paths from $v_x$ to $v_y$ for $v_y \in B_{s_t}\setminus v_x$. In Figure~\ref{treeNode}, the distances of border vertex $v_8$ from other border vertices $v_1$ and $v_5$ are 16 and 21 respectively. Thus, $d_B^{min}(v_8, G_{1_1}) = 16$.

The minimum POI distance $d_p^{min}(v_x, G_{s_t})$ of a border vertex $v_x$ is defined as is the distance from $v_x$ to its closest POI in $G_{s_t}$. In Figure~\ref{treeNode}, the distances of border vertex $v_8$ from POIs $v_1$, $v_3$ and $v_{10}$ are 16, 10
and 20 respectively. Thus, $d_p^{min}(v_8, G_{1_1})=10$.

\revised{After the $Ct$-tree construction, for each vertex $v$ in $G$, we store a pointer to each $Ct$-tree node whose subgraph contains $v$. Since the height is $s^{max}-1$, this requires adding at most $O(s^{max})$ pointers for each $v$. We remark that the height $h$ of the Ct-tree can be controlled if needed. Specifically, to ensure a height $h$, the domain of possible \revised{ESS} values is divided in $h$ contiguous intervals and, in each iteration, the edges with \revised{ESS} in the next smallest interval are removed. E.g., if \revised{ESS} domain is $1$ to $10$, a Ct-tree of height $5$ can be constructed by first removing edges with \revised{ESS} in range $(0,2]$ and then $(2,4], (4,6]$ and $(8,10]$ (in this order).}

\subsection {Query Processing}{\label{Ct-SSN}}

\ifTKDE

\else
In Section~\ref{sec:SNN_search_Ct}, we first discuss our technique for $k$ SNN search using the $Ct$-tree properties and in Section~\ref{Ct_tree_algorithm}, we present the detailed algorithms.
\fi

\subsubsection{$k$SNN search}\label{sec:SNN_search_Ct}
The efficiency of any approach for evaluating a $k$SNN query depends on the area of the graph search space for finding the safest paths having distances less than $d_c$ from $v_l$ to the POIs and the number of POIs considered for identifying $k$SNNs. The $Ct$-tree structure \revised{and Property~\ref{ss_property1_1} of our PSS measure} allow us to start the search from the smallest and safest road network subgraph that has the possibility to include $k$SNNs for $v_l$. By construction of $Ct$-tree, it is guaranteed that the subgraph of a child node is smaller and safer than that of its parent node. Thus, starting from the root node, our approach recursively traverses the child nodes that include $v_l$. The traversal ends once a child node that includes less than $k$ POIs and $v_l$ is reached. The parent of the last traversed child node is selected as the starting node of our $k$SNN search. Note that the subgraph $G_{s_t}$ of the starting node includes greater than or equal to $k$ POIs. 

If the distances of the paths from $v_l$ to at least $k$ POIs in $G_{s_t}$ is smaller than $d_c$, then our approach does not need to expand $G_{s_t}$. This is because, by definition, the edges that connect the border vertices in $G_{s_t}$ to other vertices that are not in $G_{s_t}$ have lower \revised{ESS}s than those of the edges in $G_{s_t}$. If there are less than $k$ POIs in $G_{s_t}$ whose safest paths from $v_l$ have distances less than $d_c$, our $Ct$-tree based approach recursively updates $G_{s_t}$ with the subgragh of its parent node until $k$SNNs are identified. 

To find the safest path having distance less than $d_c$ from $v_l$ to the POIs in $G_{s_t}$, we improve the INE based safest path search discussed in Section~\ref{INE} by incorporating novel pruning techniques using $Ct$-tree properties. Specifically, the minimum border distance and the minimum POI distance stored in the $Ct$-tree node allow us to develop Pruning Rules~\ref{lemma:pruning3} and~\ref{lemma:pruning4} to further refine the search space in $G_{s_t}$.

 \begin{pruning}
 \label{lemma:pruning3}
  A path $pt_{lj}$ can be pruned if $dist(pt_{lj}) +  d_B^{min}(v_j, G_{s_t}) \geq d_c$ and $dist(pt_{lj}) + d_p^{min}(v_j, G_{s_t}) \geq d_c$, where $v_j$ is a border vertex of $G_{s_t}$ and $d_B^{min}(v_j, G_{s_t})$ and $d_p^{min}(v_j, G_{s_t})$ are the minimum border distance and the minimum POI distance  of $v_j$, respectively.
\end{pruning}

\begin{pruning}
\label{lemma:pruning4}
    A path $pt_{lj}$ can be pruned if $dist(pt_{lj}) + d_B^{min}(v_j, G_{s_t}) \geq d_c$  and $dist(pt_{lj}) + d_p^{min}(v_j, G_{s_t}) \geq maxD$, where $v_j$ is a border vertex of $G_{s_t}$, $P_{s_t}$ represents the set of POIs in $G_{s_t}$, and $maxD$ represents the maximum of the current shortest distances of the POIs in $P_{s_t}$ from $v_l$, i.e., $maxD= \max_{p_i \in P_{s_t}} D_{sh}[i]$. 
\end{pruning}

\begin{figure}[hbt!]
    \centering
    \includegraphics[width=.45\textwidth]{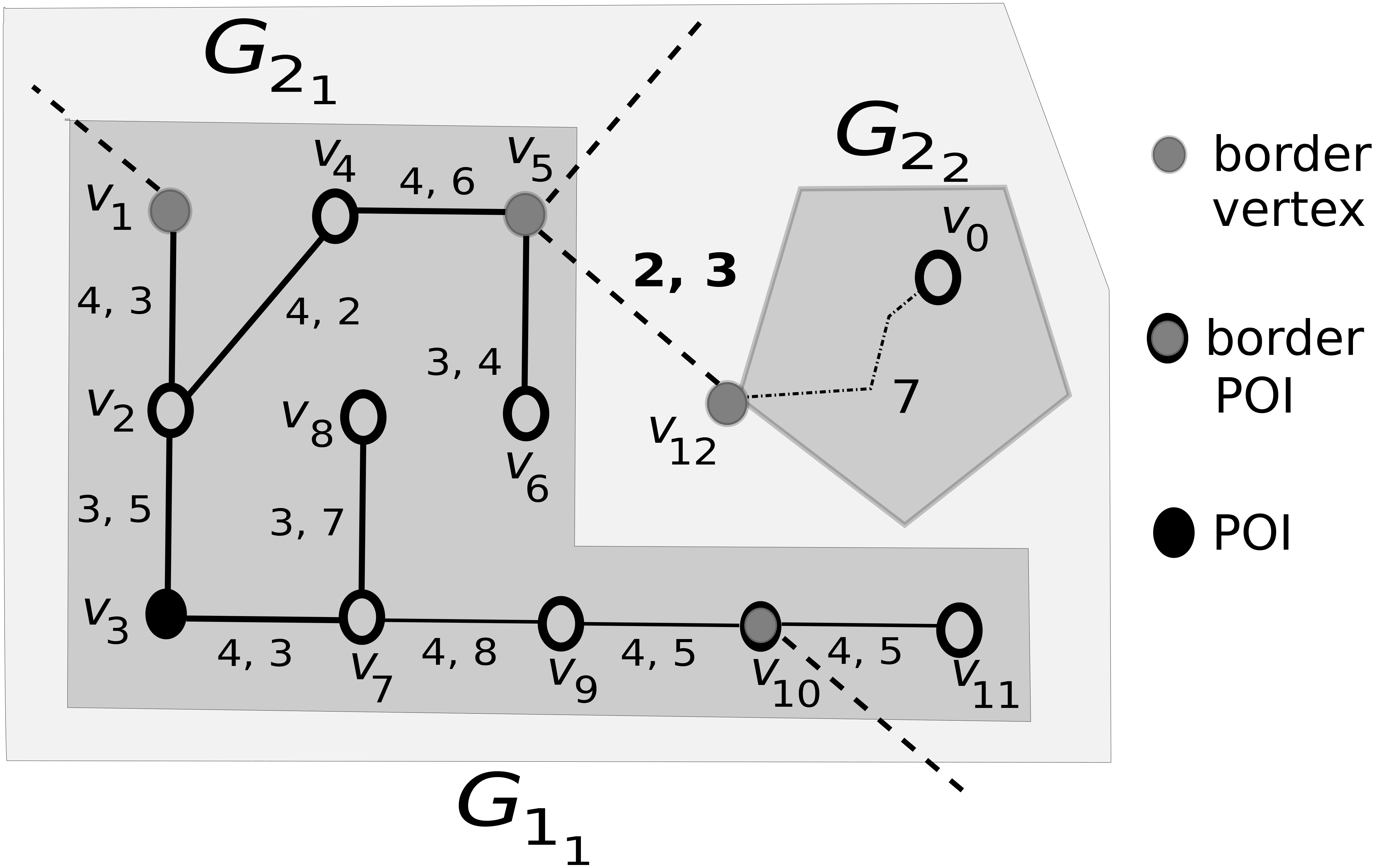}
    \caption{ $pt_{05}$ can be pruned using Pruning Rule~\ref{lemma:pruning3} for a 
    query location $v_0$ and  $d_c=20$}
    \label{fig:pruning3}
\end{figure}

If the first condition that uses the minimum border distance in Pruning Rule~\ref{lemma:pruning3} or~\ref{lemma:pruning4} becomes true then it is guaranteed that the expanded path through $pt_{lj}$ cannot cross $G_{s_t}$ to reach a POI outside of $G_{s_t}$ due to the violation of distance constraint. On the other hand, satisfying the second condition in Pruning Rule~\ref{lemma:pruning3} means that the expanded path through $pt_{lj}$ cannot reach a POI inside $G_{s_t}$ due to the violation of distance constraint. For the second condition, Pruning Rule~\ref{lemma:pruning4} exploits that if the POIs inside $G_{s_t}$ are already reached using other paths then $maxD$, the maximum of the current shortest distances of the POIs in $G_{s_t}$ can be used to prune a path. If the second condition in Pruning Rule~\ref{lemma:pruning4} becomes true then it is guaranteed that the expanded path through $pt_{lj}$ cannot provide paths that are safer than those already identified for the POIs in $G_{s_t}$ (please see Pruning Rule~\ref{lemma:pruning2} for details).

\begin{figure}[hbt!]
    \centering
    \includegraphics[width=.40\textwidth]{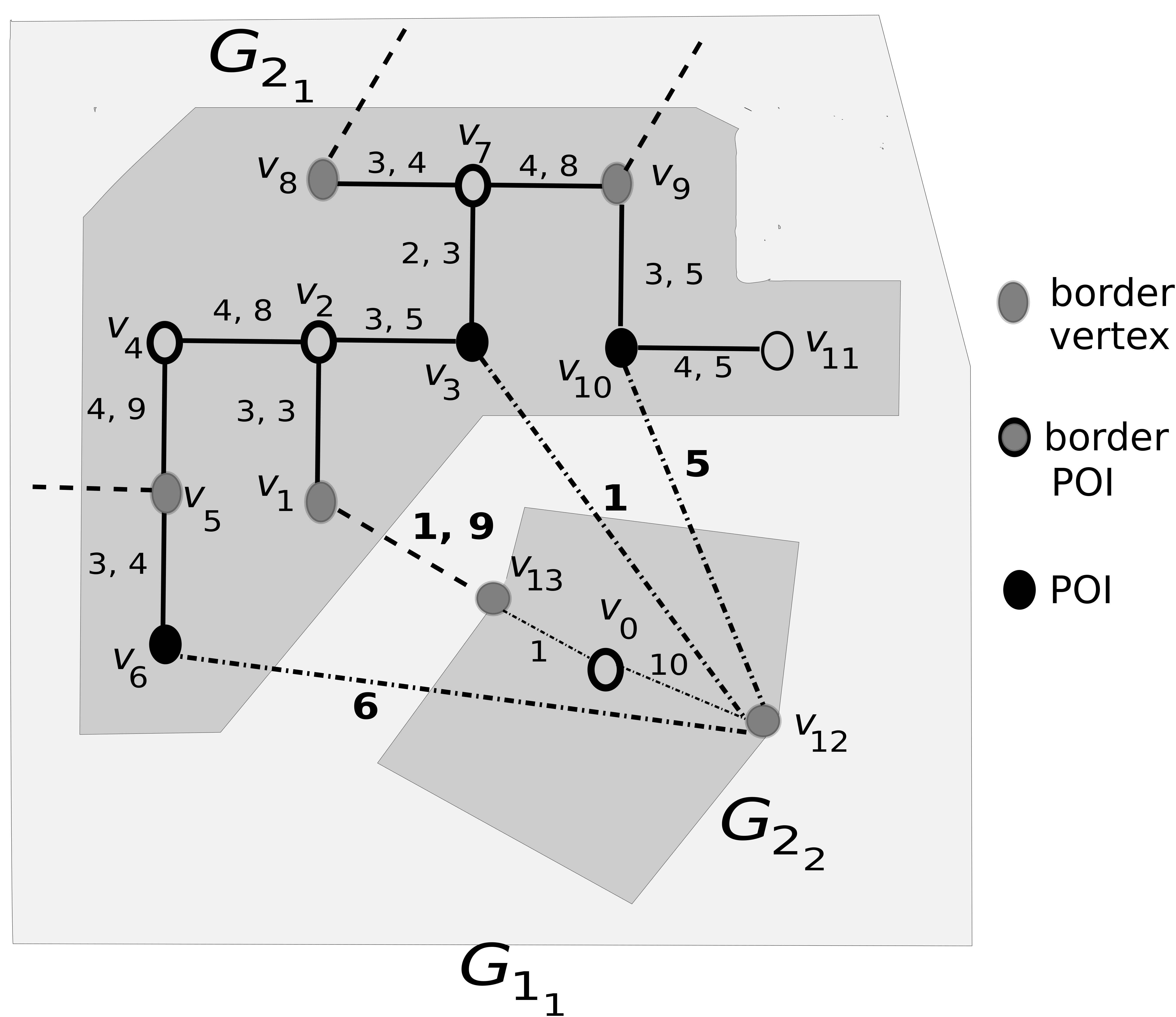}
    \caption{ $pt_{01}$ can be pruned using Pruning Rule~\ref{lemma:pruning4} for a 
    query location $v_0$ and  $d_c=20$}
    \label{fig:pruning4}
\end{figure}

Pruning Rule~\ref{lemma:pruning4} can prune more paths than Pruning Rule~\ref{lemma:pruning3} when $maxD$ is less than $d_c$, i.e., at least one path from $v_l$ to every POI in $G_{s_t}$ have been identified. On the other hand, Pruning Rule~\ref{lemma:pruning3} is better than Pruning Rule~\ref{lemma:pruning4} when $maxD$ is $\infty$, i.e., no path has yet been identified for a POI in $G_{s_t}$. Hence we consider all pruning rules (Pruning Rule~\ref{lemma:pruning1}--Pruning Rule~\ref{lemma:pruning4}) to check whether a path can be pruned.

Figure~\ref{fig:pruning3} shows an example where a path $pt_{05}$ is pruned using Pruning Rule~\ref{lemma:pruning3} for a query location $v_0$ and  $d_c=20$. In the example, $v_5$ is a border vertex of $G_{2_1}$, $dist(pt_{05})=10$, $d_B^{min}(v_5, G_{2_1})=11$ and  $d_p^{min}(v_5, G_{2_1})=13$. Here, both $dist(pt_{05})+d_B^{min}(v_5, G_{2_1})$  and $dist(pt_{05})+d_p^{min}(v_5, G_{2_1})$ are greater than $d_c$. Hence according to Pruning Rule~\ref{lemma:pruning3}, path $pt_{05}$ can be pruned. From the figure we also observe that if we expand $pt_{05}$, it cannot reach a POI outside $G_{2_1}$ through a border vertex ($v_1$ or $v_{10}$) or a POI ($v_3$ or $v_{10}$) in $G_{2_1}$ due to the violation of the distance constraint.

In Figure~\ref{fig:pruning4}, the POIs ($v_3$, $v_6$ and $v_{10}$) in $G_{2_1}$ are already reached using other paths and $D_{sh}[3]=11$, $D_{sh}[6]=16$ and $D_{sh}[10]=15$. Thus $maxD=16$. The query location is $v_0$, $d_c=20$, $dist(pt_{01})=10$, $v_1$ is a border vertex of $G_{2_1}$, $d_B^{min}(v_1, G_{2_1})= 15$ and  $d_p^{min}(v_1, G_{2_1})= 8$. Here, $dist(pt_{01})+d_B^{min}(v_1, G_{2_1})$ is greater than $d_c$ and $dist(pt_{01})+d_p^{min}(v_1, G_{2_1})$ is greater than $maxD$ but not $d_c$. Thus, in this case path $pt_{01}$ is pruned using Pruning Rule~\ref{lemma:pruning4}.

\begin{algorithm}
        \caption{$Ct$-tree-$k$SNN($v_l, k, d_c$)}
        \label{algo:GetRoute}
        \begin{algorithmic}[1]
            \STATE  $Initialize(Q_{cur}, Q_{next}, D_{sh})$
            \STATE  $G_{s_t} \gets Contains(v_l)$\label{Ct:contains}
            \STATE  $Enqueue(Q_{cur},v_l, \infty, 0)$
                \WHILE{$Q_{cur} != \emptyset$}
                    \STATE  $pt_{li}, pss(pt_{li}), dist(pt_{li}) \gets Dequeue(Q_{cur})$  %
                    \IF{ $dist(pt_{li}) < D_{sh}[i]$}
                        \STATE $D_{sh}[i]  \gets dist(pt_{li})$ 
                    \ENDIF    
                    \IF{ $isPOI(v_i)$ and $!Include(A,v_i)$}
                            \STATE $A \gets A \cup \{pt_{li}\}$
                            \IF{ $A.size() == k $}
                                \STATE \textbf{return} $A$
                            \ENDIF
                    \ENDIF
                    \IF{ $v_i \in B_{s_t}$}
                        \STATE $Enqueue(Q_{next},pt_{li}, pss(pt_{li}), dist(pt_{li}))$
                    \ENDIF
                
                   \FOR{$e_{ij} \in E_{s_t}$}
                            \IF{$!PrunePath(pt_{lj}, d_c, D_{sh}, G_{s_t})$}
                                \STATE $Enqueue(Q_{cur}, pt_{lj}, pss(pt_{lj}), dist(pt_{lj}))$ 
                           \ENDIF
                   \ENDFOR
                   \IF{$Q_{cur} = \emptyset$}
                        \STATE  $G_{s_t} \gets GetParent(G_{s_t})$
                        \STATE  $Q_{cur} \gets Q_{next}$
                        \STATE  $Q_{next} \gets \emptyset$
                   \ENDIF
                \ENDWHILE
          \STATE \textbf{return} $A$
       \end{algorithmic}
    \end{algorithm}
\subsubsection {Algorithm}
\label{Ct_tree_algorithm}

\ifTKDE
Algorithm~\ref{algo:GetRoute}, Ct-$k$SNN, shows the pseudocode to find $k$SNNs in the road network using the $Ct$-tree.
\else
Algorithm~\ref{algo:GetRoute}, Ct-$k$SNN, shows the pseudocode to find $k$ SNNs in the road network using the $Ct$-tree index structure. The inputs to the algorithm are $v_l$, $k$ and $d_c$. The algorithm returns $A$, an array of $k$ entries, where each entry includes the safest path from $v_l$ to a POI whose distance is less than $d_c$. The entries in $A$ are sorted in the descending order based on the PSS of the safest path.
\fi

The algorithm uses two priority queues $Q_{cur}$ and $Q_{next}$, where $Q_{cur}$ is used for the path expansion in $G_{s_t}$, and $Q_{next}$ stores the paths that will be expanded WHILE searching the parent subgraph of $G_{s_t}$. Each entry of these queues represents a path, the PSS and the path's length. The entries in a queue are ordered in descending order based on their PSSs. The algorithm uses an array $D_{sh}$, where $D_{sh}[i]$ stores the distance of the shortest dequeued path from $v_l$ to $v_i$.

The algorithm starts with initializing two priority queues $Q_{cur}$ and $Q_{next}$ to $\emptyset$ and $D_{sh}[i]$ for each $v_i$ to $d_c$ using Function $Initialize$ (Line 1). \revised{Then, starting from the root node, Function $Contains$ recursively traverses the child nodes using the stored pointers of $v_l$ until it identifies the smallest subgraph $G_{s_t}$ of a $Ct$-tree node that includes $v_l$ and has at least $k$ POIs (Line 2).} The  $v_l$ and the corresponding path information are enqueued to $Q_{cur}$. The algorithm iteratively processes the entries in $Q_{cur}$ until it becomes empty or $k$SNNs are  found (Lines 4--28).

In every iteration, the algorithm dequeues a path $pt_{li}$ from $Q_{cur}$ and updates  $D_{sh}[i]$ of the last vertex $v_i$ of the dequeued path if $dist(pt_{li}) < D_{sh}[i]$ (Lines 5--8). If $v_i$ represents a POI that is not already present in $A$, then $pt_{li}$ is added to $A$, and if $A$ includes $k$ entries, the answer is returned (Lines 9--14). 

If $v_i$ is a border vertex of $G_{s_t}$, then there is at least one outgoing edge from $v_i$ with safety score smaller 
than or equal to $s$, which might need to be later considered if $k$SNNs are not found in current $G_{s_t}$. Thus, $pt_{li}$ and the corresponding path information are enqueued to $Q_{next}$ (Lines 15--17).  Note that the \revised{ESS} of $G_{s_t}$ are 
larger than $s$.

Next, for each outgoing edge $e_{ij}$ of $v_i$ in $G_{s_t}$, the algorithm checks whether the newly formed path $pt_{lj}$ by adding $e_{ij}$ at the end of $pt_{li}$ can be pruned using $PrunePath$ function ($PrunePath$ is elaborated below). If the path is not pruned, then $pt_{lj}$ and the corresponding path information are enqueued to $Q_{cur}$ (Lines 18--22).

At the end of the iteration, the algorithm checks whether the exploration of $G_{s_t}$ is complete, i.e., $Q_{cur}$ is empty. If this condition is true, then it means that the safest paths having distances less than $d_c$ from $v_l$ to $k$SNNs are not included in $G_{s_t}$. Thus, the algorithm sets the parent of $G_{s_t}$ as $G_{s_t}$, assigns $Q_{next}$ to $Q_{cur}$ and resets $Q_{next}$ to $\emptyset$ (Lines 23-27).   

\textbf{\emph{PrunePath}}. Algorithm~\ref{algo:pruning} returns $true$ if path $pt_{ij}$ can be pruned by any of our pruning criteria and $false$ otherwise. The algorithm first checks 
whether $pt_{ij}$ can be pruned using the criteria in Pruning Rules~\ref{lemma:pruning1} or~\ref{lemma:pruning2}. Note that an entry $D_{sh}[x]$ for a vertex $v_x$ is initialized to $d_c$ and later it gets updated once a path to $v_x$ is dequeued from the queue (see Line 7 in Algorithm~1).

If $pt_{ij}$ is not pruned, Function $checkBorder(G_{s_t},v_j)$ checks whether $v_j$ is a border vertex of $G_{s_t}$ or one of its descendants. If so, it returns $true$ for $IsBorder$ and the node $G_{{s^\prime}_{t^\prime}}$ for which $v_j$ is a border vertex. Otherwise, $IsBorder$ is set to $false$. By construction of the $Ct$-tree, a border vertex of a $Ct$-tree node is also the border vertex of its descendants. Thus, there may be multiple nodes for which $v_j$ is a border vertex, in which case, $G_{{s^\prime}_{t^\prime}}$ is chosen to be the highest node in the $Ct$-tree for which $v_j$ is a border vertex.

If $v_j$ is a border vertex then the algorithm checks whether $pt_{ij}$ can be pruned using Pruning Rules~\ref{lemma:pruning3} or~\ref{lemma:pruning4} (Lines 5--8). One of the pruning condition that uses the minimum border distance is same in both Pruning Rules~\ref{lemma:pruning3} and~\ref{lemma:pruning4}, which is checked in Line 5. The left part of the other condition, adding the minimum POI distance with $dist(pt_{ij})$ is also same in both pruning rules. The right part is $d_c$ for Pruning Rule~\ref{lemma:pruning3} and $maxD$ for Pruning Rule~\ref{lemma:pruning4}, where $maxD$ represents the maximum of the current shortest distances of $v_l$ to the POIs in $G_{{s^\prime}_{t^\prime}}$. The shortest distance of every vertex (including POIs) is initialized to $d_c$ (Line 1 of Algorithm~\ref{algo:GetRoute}). Thus, $maxD$ is initially $d_c$ and later may become less than $d_c$ when the paths to POIs in $G_{{s^\prime}_{t^\prime}}$ are dequeued from the priority queue (Lines 5--7 of Algorithm~\ref{algo:GetRoute}).

Since the minimum border distance and minimum POI distance of a $Ct$-tree node are greater than or equal to those of its descendants, it is not required to check these pruning conditions for $v_j$ for the descendants of $G_{{s^\prime}_{t^\prime}}$ separately.

\begin{algorithm}[htbp]
\caption{$PrunePath(pt_{lj}, d_c, D_{sh}, G_{s_t})$}
\label{algo:pruning}
\begin{algorithmic}[1]
            \IF{$dist(pt_{lj})\geq D_{sh}[j]$}
                \STATE \textbf{return} true 
            \ENDIF
            
            \STATE $isBorder{G_{{s^\prime}_{t^\prime}}} \gets  checkBorder(G_{s_t}, v_j)$
            \IF{$isBorder == true$ and $dist(pt_{lj})+d_B^{min}(v_j, G_{{s^\prime}_{t^\prime}}) \geq d_c$}
                \IF{$dist(pt_{lj})+d_p^{min}(v_j, G_{{s^\prime}_{t^\prime}}) \geq maxD$}
                     \RETURN true
                \ENDIF
            \ENDIF
            \RETURN false
        \end{algorithmic}
    \end{algorithm}

\ifTKDE
\revised{In full version~\cite{biswas2021safest} of this paper, we provide complexity analysis of the algorithm and discuss how to update $Ct$-tree when there are updates to the road network and POIs.}
\else
\revised{\emph{Complexity Analysis.}  Since the height $h$ of the $Ct$-tree is bounded by $s^{max}-1$, the $Contains$ function (Line~\ref{Ct:contains}, Algorithm~\ref{algo:GetRoute}) to obtain intial $G_{s_t}$ takes $O(s^{max})$, where $s^{max}$ is typically a small constant (e.g., 5, 10 or 15). To expand search range and update $G_{s_t}$, Function $GetParent$ is called at most $(s^{max}-1)$ times and thus, its overall worst case time complexity is also $O(s^{max})$.}

\revised{
Let the total number of valid paths from $v_l$ to any node in $G_{s_t}$ be $n^\prime_p$. Then the worst case time complexity for finding $k$ SNNs by applying the INE based search using a priority queue is $O(n^\prime_p \log(n^\prime_p))$. The number of total paths reduces when we apply Pruning Rules~\ref{lemma:pruning2},~\ref{lemma:pruning3}, and~\ref{lemma:pruning4}. Let $r^\prime$ be the combined effect factor of these three pruning rules. Thus, by applying the pruning rules, the worst case time complexity of the INE based search in $G_{s_t}$ becomes $O(\frac{n^\prime_p}{r^\prime} \log(\frac{n^\prime_p}{r^\prime}))$.  
Thus, the worst case time complexity for the $Ct$-tree based approach is $O(s^{max} +\frac{n^\prime_p}{r^\prime}\log(\frac{n^\prime_p}{r^\prime}))$. Note that this complexity is significantly better than the worst case time complexity of INE-based approach because $s^{max}$ is typically a small constant, $n^\prime_p/r^\prime$ is significantly smaller than $n_p/r$ because $G_{s_t} \subseteq G$ and $Ct$-tree uses additional pruning rules.}

\subsection {Ct-tree Update}{\label{Ct_update}}
A $Ct$-tree needs an update when there is any change in the edges or POIs of the road network. 

\emph{\textbf{Adding/removing an edge or change in \revised{ESS}}} 
When an edge $e_{xy}$ with \revised{ESS} $s$ is added/removed, the $Ct$-tree nodes whose subgraphs allow \revised{ESS} $s$ and include $v_x$ and/or $v_y$ may get affected. If $v_x$ and $v_y$ are located at different subgraphs then two subgraphs become connected when $e_{xy}$ is added and their corresponding $Ct$-tree nodes are merged into one. On the other hand, the removal of $e_{xy}$ from a subgraph may divide the subgraph into two components and the corresponding $Ct$-tree node to two nodes. This process recursively continues by checking the child nodes that include $v_x$ and/or $v_y$ until the subgraphs whose allowed \revised{ESS} is greater than $s$ is reached. 

When an \revised{ESS} changes, the \revised{ESS} is simply updated in the subgraphs where the edge already exists. If an \revised{ESS} increases (e.g., \revised{ESS} increases from 3 to 4), the edge is added to the subgraphs, where the new \revised{ESS} is allowed. On the other hand, If an \revised{ESS} decreases (e.g., \revised{ESS} decreases to 3 from 4), the edge is removed from the subgraphs where the new \revised{ESS} is not allowed. The addition or removal of an edge may result in the merge or division of the subgraph(s) and their corresponding $Ct$-tree node(s).

Although the travel time associated with a road represented by an edge may change frequently, the changes in \revised{ESS} values are not frequent, e.g., the \revised{ESS} may decrease when there is a reported crime on the edge. Also, in most real world applications, the updates to \revised{ESS}s are periodic (e.g., all \revised{ESS} values may be updated at the end of every week based on the recent crime data). Even if an \revised{ESS} changes, the change normally happens in small step (e.g., 5 to 4). It is an uncommon scenario that an \revised{ESS} suddenly changes from 5 to 1. As a result the update process due to an \revised{ESS} change affect only a small portion of the $Ct$-tree structure and incurs low processing overhead.

\emph{\textbf{Adding/removing a POI:}} When a POI is added or removed, the information on the number of POIs stored in the $Ct$-tree nodes are updated accordingly.

Irrespective of the $Ct$-tree structure update, adding/removing an edge $e_{xy}$ or a change in the \revised{ESS} of $e_{xy}$ or adding/removing a POI may require update in the border vertex set, the minimum border distance, the minimum POI distance of border vertices for the $Ct$-tree nodes whose subgraphs include $v_x$ and/or $v_y$ \revised{and the stored pointers to the $Ct$-tree nodes for the vertices in the road network graph.}

\fi

%% file: sections/6_VoronoiBasedSolution.tex
\section{Safety Score Based Network Voronoi Diagram}
{\label{voronoiBasedSolution}}

The concept of the network Voronoi diagram (NVD)~\cite{OkabeBSCK00, OkabeSFSO08} has been shown as an effective method for faster processing of $k$ nearest neighbor queries~\cite{KolahdouzanS04,SafarET09}. The traditional NVD is computed based on  distance  and does not apply for processing $k$SNN queries. In Section~\ref{sec:SNVD}, we introduce \emph{safety score based network Voronoi diagram (SNVD)} and present its properties.  
\ifTKDE
We develop an efficient technique to compute the $k$SNNs using the SNVD. Our solution is composed of two phases: preprocessing (Section~\ref{phase1}) and query processing (Section~\ref{phase2}). 
\else

We develop an efficient technique to compute the $k$SNNs using the SNVD. Our solution is composed of two phases: preprocessing  and query processing. 
In Sections~\ref{phase1} and~\ref{phase2}, we elaborate the preprocessing steps and the query processing algorithm, respectively. 
In Section~\ref{SNVD_update}, we discuss how to update the SNVD for a change in the road network. 
\fi

\begin{figure}[htbp]
   \begin{center}
        \begin{tabular}{cc}
            \hspace{-5mm}
            \resizebox{40mm}{30mm}{\includegraphics{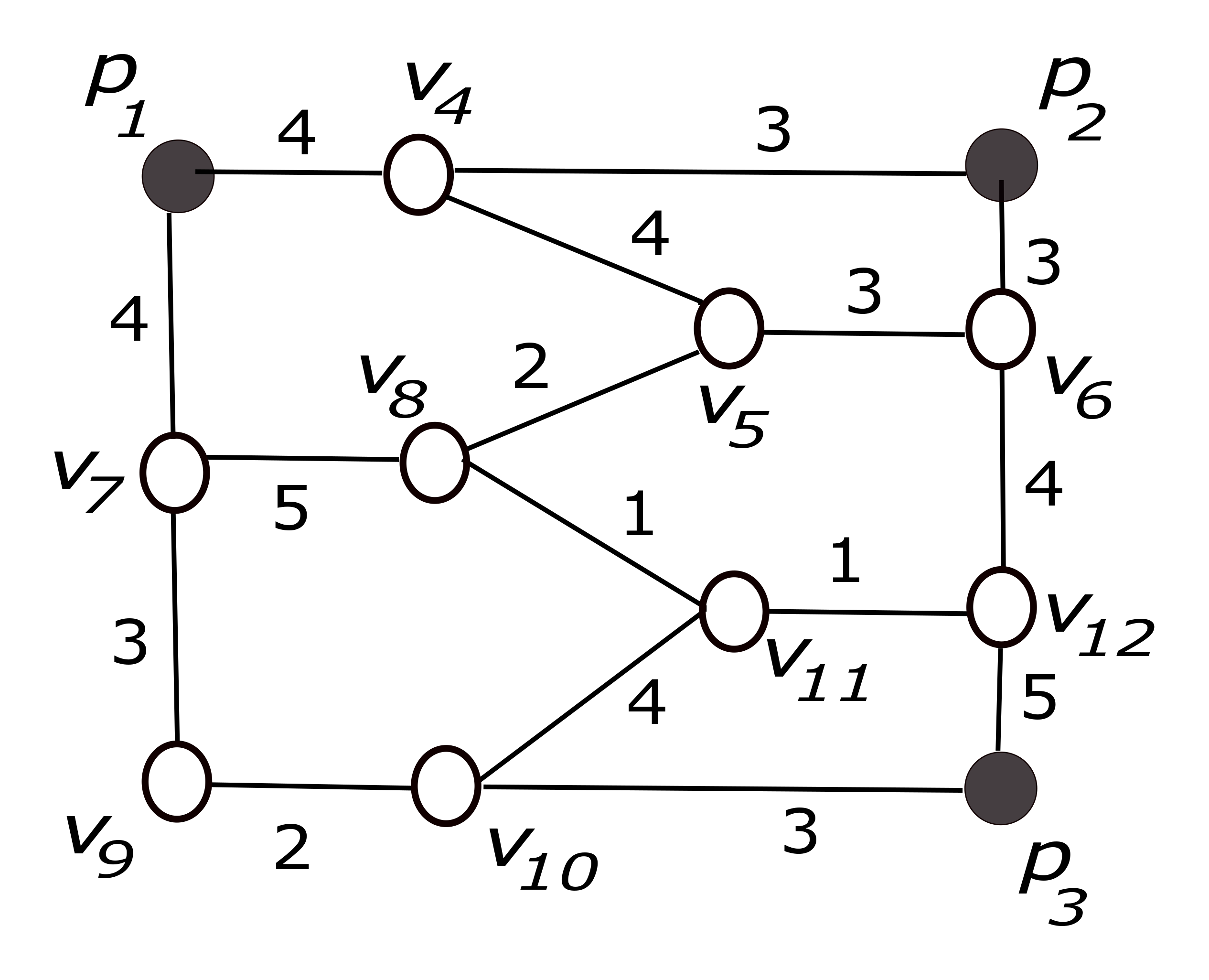}} &
            \hspace{-6mm}
            \resizebox{40mm}{30mm}{\includegraphics{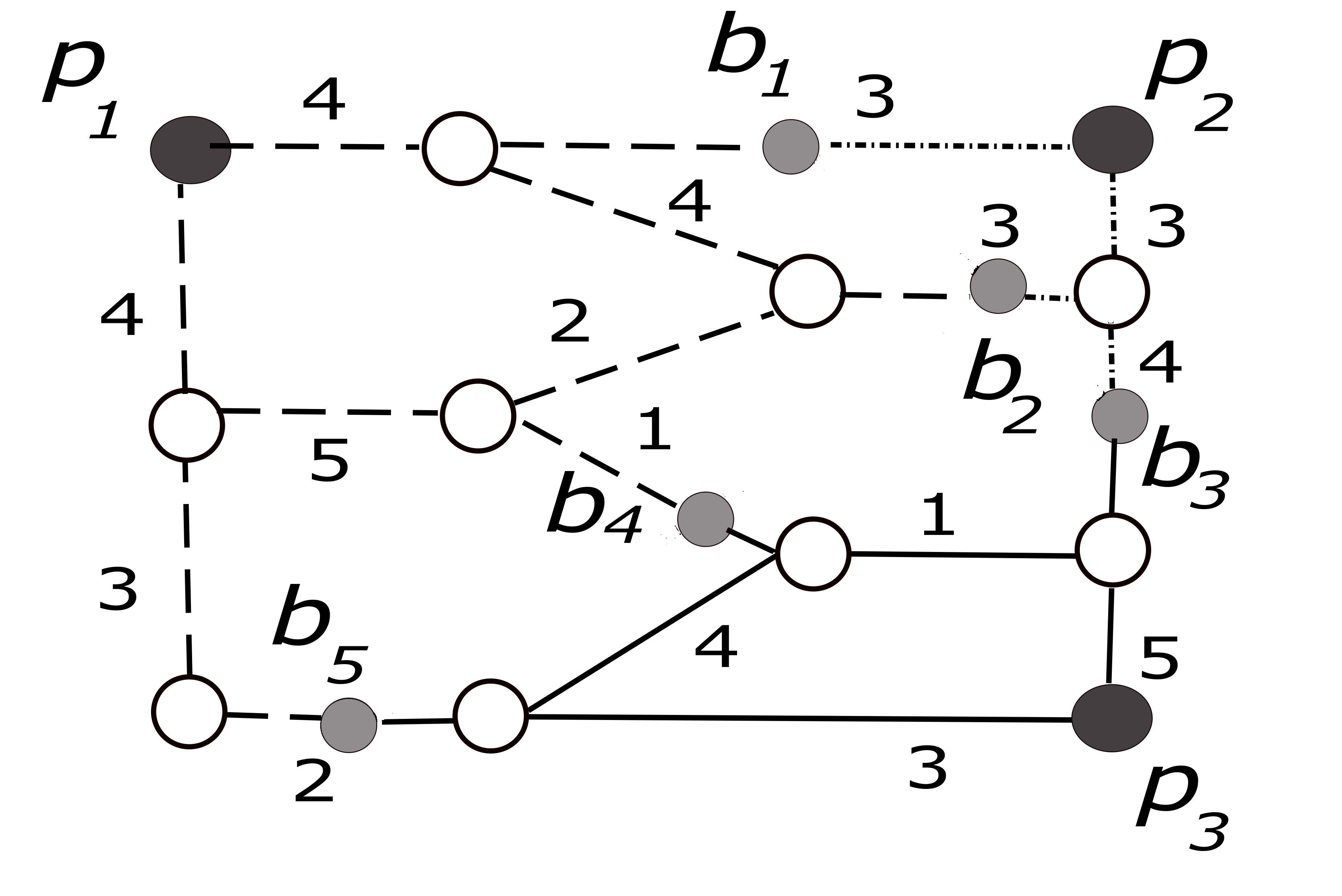}}\\ 
            \hspace{1mm} \scriptsize{(a) Original Graph \textsc{}\hspace{2mm}}  &  \scriptsize{(b) SNVD \textsc{}\hspace{0mm}}\\
        \end{tabular}
        \caption{An example of SNVD}
        \label{fig:SNVD_example}
    \end{center}

\end{figure}

\subsection{SNVD Properties}
\label{sec:SNVD}

An SNVD divides the road network graph into subgraphs such that each subgraph contains a single POI which is the \emph{unconstrained safest neighbor} for every point of the subgraph. We differentiate the unconstrained safest neighbor from the safest nearby neighbor (SNN) by considering \emph{unconstrained safest path}. For an unconstrained safest path, we assume a sufficiently large value for $d_c$ such that every simple path in the graph has length less than $d_c$, e.g., $d_c$ is just bigger than the sum of the lengths of all edges in the graph. The subgraphs are called Voronoi cells and a point that defines the boundary between two adjacent Voronoi cells is a called a border vertex. The Voronoi cell of POI $p_i$ is denoted as ${VC}_i$ and the border vertex set for ${VC}_i$ is denoted as $B_i$.

Figure~\ref{fig:SNVD_example} shows an example of an SNVD. Black vertices represent the generator POIs $p_1$, $p_2$ and $p_3$, white vertices represent the road intersections, and grey vertices represent the border vertices.

\begin{figure*}[htbp]
   \begin{center}
         \resizebox{150mm}{!}{\includegraphics{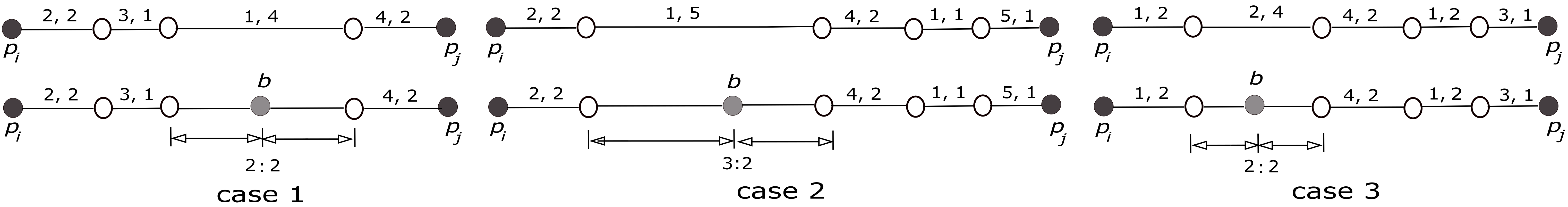}} 
         \caption{Example border vertices of an SNVD, weights associated with each edge are $w_{ij}^{ss}$, $w_{ij}^d$,  respectively}
        \label{fig:Voronoi_boders}
    \end{center}
\end{figure*}

Unlike traditional distance based NVD~\cite{SafarET09, XuanZTSSG09}, a border vertex $b$ between two adjacent Voronoi cells ${VC}_i$ and ${VC}_j$ may not have equal PSS for its unconstrained safest paths to $p_i$ and $p_j$. Figure~\ref{fig:Voronoi_boders} shows examples of such scenarios. When the PSSs for $b$'s unconstrained safest paths to $p_i$ and $p_j$ differ, the unconstrained safest neighbor of $b$ is one of the POIs $p_i$ and $p_j$ for which $b$ has the higher PSS. 

The following lemmas shows the properties of SNVD that we exploit in our solution.

\begin{lemma}
\label{SNVD_p1}
Let $p_1, p_2, \dots, p_{k-1}$ represent the $k-1$ unconstrained safest POIs of $v_l$, respectively. The unconstrained safest path from $v_l$ to its $k^{th}$ unconstrained safest POI $p_k$ can only go through $\{{VC}_1, {VC}_2, \dots, {VC}_{k-1}, {VC}_k\}$.
\end{lemma}
\begin{proof}
Assume that the unconstrained safest path from $v_l$ to $p_k$ goes through a vertex $v$ of ${VC}_t$, where $t \notin \{1,2,\dots, k\}$. By construction of SNVD, POI $p_t$ is safer than POI $p_k$ for $v$. Since the unconstrained safest path from $v_l$ to $p_k$ goes through $v$, POI $p_t$ is also safer than POI $p_k$ for $v_l$. Thus $t \in \{1,2,\dots, k\}$, which contradicts our assumption.
\qed
\end{proof}

\begin{lemma}
\label{SNVD_p2}
Let $v_1, v_2, \dots, v_{k-1}$ represent the $k-1$ unconstrained safest POIs of $v_l$, respectively. The $k^{th}$ unconstrained safest POI $p_k$ of $v_l$ lies in one of the adjacent Voronoi cells of ${VC}_1, {VC}_2, \dots, {VC}_{k-1}$.
\end{lemma}

\begin{proof}
If the $k^{th}$ unconstrained safest POI $v_k$ of $v_l$ does not lie in one of the adjacent Voronoi cells of ${VC}_1, {VC}_2, \dots, {VC}_{k-1}$, then the unconstrained safest path between $v_l$ and $v_k$ must go through a vertex which does not lie in ${VC}_1, {VC}_2, \dots, {VC}_{k-1}, {VC}_k$, which contradicts Lemma~\ref{SNVD_p1}. Thus, the $k^{th}$ unconstrained safest POI of $v_l$ lies in one of the adjacent Voronoi cells of ${VC}_1, {VC}_2, \dots, {VC}_{k-1}$.
\qed
\end{proof}

\subsection{Prepossessing} \label{phase1}

The following information will be precomputed and stored for the faster processing of $k$SNN queries:
\begin{itemize}
\item Voronoi cells, where each Voronoi cell ${VC}_i$ has the POI $p_i$ as the generator, a set $B_i$ of border vertices, and a list $L_i$ of pointers to adjacent Voronoi Cells
\item For every vertex $v\in B_i$, the unconstrained safest path between $v$ and $p_i$ and its PSS
\item For every pair of border vertices $v_i\in B_i$ and $v_j\in B_i$, the unconstrained safest path between $v_i$ and $v_j$ that lies completely inside $VC_i$ and its PSS
\item For every $v_x \in B_i$  the minimum border distance and the minimum POI distance, where the minimum border distance of a border vertex is defined as $argmin_{v_y \in B_i \setminus v_x}dist(pt^{sh}_{xy})$ and the minimum POI distance is $dist(pt^{sh}_{xi})$ and the shortest paths for the minimum border distance and the minimum POI distance are computed by considering only the subgraph represented by Voronoi cell ${VC}_i$
\item \revised{For every vertex $v$ in the road network graph $G$, a pointer to the Voronoi cell whose POI generator is the unconstrained safest neighbour of $v$}
\end{itemize}

To construct the SNVD, we first apply parallel Dijkstra algorithm~\cite{Erwig00} to find the unconstrained safest POI for each vertex in the road network. Starting from a start vertex, the Dijkstra algorithm keeps track of the unconstrained safest path for every visited vertex. When the distance for the safest path is unconstrained, no path gets pruned and thus unlike INE, it is sufficient to only expand the unconstrained safest path from a vertex.

Then we compute the border vertices as follows: (i) if the unconstrained safest paths from $v$ to two POIs $p_i$ and $p_j$ have the same PSS, then $v$ is considered as a border vertex of Voronoi cells ${VC}_i$ and ${VC}_j$, and (ii) if two end vertices of an edge have different unconstrained safest POIs, say $p_i$ and $p_j$, and the \revised{ESS} associated with the edge is $s$, then the point $b$ of the edge that provides equal distance associated with $s$ in both of the $b$'s safest paths to $p_i$ and $p_j$ is identified as a border vertex of Voronoi cells ${VC}_i$ and ${VC}_j$. Figure~\ref{fig:Voronoi_boders} shows three example scenarios. In case 1, the end vertices of the edge with \revised{ESS} 1 has different unconstrained safest POIs and the distance of the edge is 4. There is no other edge in $pt_{ij}$ with \revised{ESS} 1. Thus $b$ is placed at the midpoint of the edge so that both of the $b$'s safest paths to $p_i$ and $p_j$ have equal distance for \revised{ESS} 1. For case 2, $b$ is not the midpoint of the edge as there is another edge in $pt_{ij}$ with \revised{ESS} 1. Note that in case 3, $b$ is placed at the mid of the edge with \revised{ESS} 2, which is not the minimum \revised{ESS} in $pt_{ij}$. This is because both of the $b$'s safest paths to $p_i$ and $p_j$ already have an edge with the same distance for \revised{ESS} 1.  

The resulted distances of the splitted edges with a border vertex $b$ may become less than 1. This occurs when the length of the edge that is splitted by $b$ is 1 and no other edge in $pt_{ij}$ has the same ESS as the splitted edge. Our PSS measure requires the edge weight to be greater than or equal to 1. Thus, after the construction of the whole SNVD, if the distance of any splitted edge becomes less than 1, we scale up the distances of all edges in the road network to ensure that the distances of the splitted edges are greater than or equal to 1 and recompute the SNVD.

To compute the unconstrained safest path between $v_i$ and $v_j$ that lies completely inside $VC_i$ for every pair of border vertices $v_i\in B_i$ and $v_j\in B_i$, we apply Dijkstra algorithm within $VC_i$.

\subsection {Query Processing}{\label{phase2}}

According to the construction of SNVD, a POI $p_i$ is the unconstrained safest neighbor for every location $v$ in a Voronoi cell $VC_i$. However, an unconstrained safest neighbor is not necessarily the SNN for a location $v$ if the distance of the unconstrained safest path from $v$ to $p_i$ is greater than or equal to the distance constraint $d_c$ specified in the query. Hence, unlike the existing NVD based nearest neighbor search algorithms~\cite{KolahdouzanS04}, we cannot simply return the POI of the Voronoi cell that includes the query location as the SNN. Since $d_c$ is a query specific parameter, it is also not possible to consider $d_c$ while computing the Voronoi cells of the SNVD. In Sections~\ref{sec:SNN_search}, we discuss our SNVD based solution to find $k$SNNs and in Section~\ref{SNVD_safestpath}, we elaborate our efficient (unconstrained and constrained) safest path computation techniques by exploiting SNVD properties.

\subsubsection{kSNN search}
\label{sec:SNN_search}
Our SNN search technique using SNVD incrementally finds the unconstrained safest neighbors and adds each of these neighbors in a candidate set. The incremental search stops when the candidate set contains at least $k$ POIs that have unconstrained safest paths with length less than $d_c$. Let $k+m$ be the size of the candidate set where $m\geq 0$. The k SNNs are guaranteed to be among the candidate set. Then, we compute the PSSs of the safest paths from $v_l$ to the remaining $m$ POIs in the candidate set to determine the query answer. The POIs whose safest paths from $v_l$ among $k+m$ candidates have $k$ largest PSSs and have distances less than $d_c$ are identified as $k$SNNs. 

Algorithm~\ref{algo:FindSSN_SNVD} shows the steps to find $k$SNNs in the road network using the SNVD. The inputs to the algorithm are $v_l$, $k$ and $d_c$. The algorithm initializes $it$ and $j$ to 1, finds the first unconstrained safest neighbor using the stored pointer (Function $FindUSN$) and adds it to the candidate POI set $C_P$ (Lines 1--2). Then the algorithm iterates in a loop until $C_P$ includes $k$SNNs. In every iteration, the algorithm checks whether the distance of the unconstrained safest path from $v_l$ to the $j^{th}$ unconstrained safest neighbor is less than $d_c$. If yes, Function $IsConstrained$ returns true and $it$ is incremented by 1 (Lines 4--6)). At the end of the iteration, the algorithm increments $j$ by 1 and finds the next ($j^{th}$) unconstrained safest neighbor using Function $NextUCN$ (Lines 7--8). When the loop ends, $FindA$ identifies the answer of the $k$SNN query in $A$ (Line 10). 

\begin{algorithm}[htbp]
\caption{SNVD-$k$SNN($v_l, k, d_c$)}
\label{algo:FindSSN_SNVD}
\begin{algorithmic}[1]
    \STATE $it, j\gets 1$
    \STATE $C_P \gets FindUSN(v_l)$
    \WHILE{$it \leq k$}
        \IF{$IsConstrained(C_P, j, d_c)$}
            \STATE $it \gets it+1$
        \ENDIF
        \STATE $j \gets j+1$         
        \STATE $C_P \gets C_P \cup NextUSN(v_l, j)$
    \ENDWHILE
    \STATE $A \gets FindA(A, C_P)$
    \RETURN $A$
\end{algorithmic}
\end{algorithm}

\textbf{\emph{Incremental computation of unconstrained safest neighbors.}} Function $FindUSN$ locates Voronoi cell ${VC}_q$ that includes $v_l$ and returns the corresponding POI $p_q$ as the first unconstrained safest neighbor. ${VC}_q$ becomes the first candidate Voronoi cell of the set $C_{VC}$. To compute the second unconstrained safest neighbor, Function $NextUSN$ adds the adjacent Voronoi cells of ${VC}_q$ to $C_{VC}$ as they include the second unconstrained safest POI of $v_l$ (Lemma~\ref{SNVD_p2}). The POI of one of these candidate Voronoi cells in $C_{VC}$ whose unconstrained safest path to $v_l$ has the second best PSS is identified as the second unconstrained safest neighbor of $v_l$ (please see Section~\ref{SNVD_safestpath} for unconstrained safest path computation technique). Similarly, to find the $j^{th}$ unconstrained safest neighbor of $v_l$, $NextUSN$ adds the adjacent Voronoi cells of the $(j-1)^{th}$ unconstrained safest neighbor's Voronoi cell in $C_{VC}$, if they are not already included in $C_{VC}$. The $j^{th}$ unconstrained safest neighbor of $v_l$ is the POI that has the $j^{th}$ best PSS for the unconstrained safest path from $v_l$ among the POIs of the candidate Voronoi cells.

\subsubsection {Safest Path Computation}{\label{SNVD_safestpath}}
Our PSS measure and the property shown in Lemma~\ref{SNVD_p1} allow us to adapt the shortest path computation technique proposed in~\cite{KolahdouzanS04} for finding the unconstrained safest paths from $v_l$ to a POI of the candidate Voronoi cell in $C_{VC}$. \revised{After $FindUSN$ locates Voronoi cell ${VC}_q$ that includes $v_l$, it computes the unconstrained safest paths from $v_l$ to $p_q$ and the border vertices of $VC_q$ using Dijkstra algorithm.} Then similar to~\cite{KolahdouzanS04}, we use precomputed unconstrained safest paths between border vertices for this purpose and reduce the processing overhead significantly. In experiments, we show that the number of border vertices of an SNVD is small with respect to the total number of vertices and thus, the overhead for storing the unconstrained safest paths for the border vertices is negligible.   

On the other hand, we cannot use precomputed unconstrained safest paths and adapt~\cite{KolahdouzanS04} to compute the safest path between $v_l$ and a POI while identifying $k$SNNs from candidate $k+m$ POIs. We improve the INE based safest path search technique discussed in Section~\ref{INE} by incorporating novel pruning techniques for search space refinement using SNVD properties and apply it to compute the safest paths that have distances less than $d_c$. Pruning Rules~\ref{lemma:pruning3_1_1} and~\ref{lemma:pruning3_1_2} use the minimum border distances and minimum POI distances of the border vertices, respectively, in the pruning condition:

\begin{pruning}
 \label{lemma:pruning3_1_1}
    Let the safest path between $v_l$ and POI $p_y$ needs to be determined. A path $pt_{lj}$ can be pruned if $dist(pt_{lj}) + d_B^{min}(v_j, {VC}_x)  \geq d_c$, where $v_j$ is a border vertex of Voronoi cell ${VC}_x$, $d_B^{min}(v_j, {VC}_x)$ represents the minimum border distance of $v_j$ and $d_c$ represent the distance constraint.
\end{pruning}

\begin{pruning}
 \label{lemma:pruning3_1_2}
    Let the safest path between $v_l$ and POI $p_x$ needs to be determined. A path $pt_{lj}$ can be pruned if $dist(pt_{lj}) + d_p^{min}(v_j, {VC}_x)  \geq d_c$, where $v_j$ is a border vertex of Voronoi cell ${VC}_x$, $d_p^{min}(v_j, {VC}_x)$ represents the minimum POI distance of $v_j$ and $d_c$ represent the distance constraint.
\end{pruning}

We have $s_k$, the upper bound of the PSS of the safest path between $v_l$ and its $k^{th}$ SNN, once $k$ unconstrained safest neighbors of $v_l$ that have distances less than $d_c$ from $v_l$ are identified. Later if a safest path from $v_l$ to a new POI is identified that has the distance less than $d_c$ and PSS higher than current $s_k$, then $s_k$ is updated to the new PSS. By exploiting the property that the PSS of a subpath is higher than that of the path (Property~\ref{ss_property1}) and $s_k$, we develop a new pruning technique as follows:

\begin{pruning}
\label{lemma:pruning5}
A path $pt_{lj}$ can be pruned if $pss(pt_{lj}) < s_k$, where $s_k$ represent the upper bound of the PSS of the safest path between $v_l$ and its $k^{th}$ SNN. \end{pruning}

Thus, in addition to Pruning Rules~\ref{lemma:pruning1} and~\ref{lemma:pruning2}, our SNVD based solution checks whether a path can be pruned using  Pruning Rules~\ref{lemma:pruning3_1_1},~\ref{lemma:pruning3_1_2}  and~\ref{lemma:pruning5} to find the safest paths that have distances less than $d_c$.

\ifTKDE
\revised{In full version~\cite{biswas2021safest} of this paper, we provide complexity analysis of the algorithm and discuss how to update SNVD when there are updates to the road network and POIs.}
\else

\revised{\textit{Complexity Analysis.}  Function $FindUSN$  identifies the unconstrained safest neighbor $p_q$ of $v_l$ in $O(1)$ using the pointer for each vertex storing its Voronoi generator. $FindUSN$ also computes the unconstrained safest paths from $v_l$ to $p_q$ and the border vertices of $VC_q$ using Dijkstra algorithm with the worst case time complexity $O(E_q\log(V_q))$ where $E_q$ is the number of edges in $VC_q$. 
The loop (Lines 3--9) iterates $(k+m)$ times. In the $j^{th}$ iteration, Function $isConstrained$ checks whether the distance of the unconstrained safest path from $v_l$ to the $j^{th}$ unconstrained safest nearest neighbour is smaller than $d_c$ in constant time. Function $NextUSN$ incrementally finds $(j+1)^{th}$ unconstrained safest neighbor of $v_l$ using precomputed distances and stored lists of neighbor SNVD cells. Thus, the overall time complexity of the loop for $(k+m)$ iterations is $O(k+m)$. The parameter $m$ is typically a small value (e.g., around 8 in our experimental study).}

\revised{Function $FindA$ needs to compute the safest paths that have distances less than $d_c$ from $v_l$ to $m$ POIs. Let $n^{\prime\prime}_p$ be the total number of valid paths from $v_l$ to any node in the Voronoi cells of the $(k+m)$ POIs. Then, the worst case time complexity for finding these safest paths by applying the INE based search using a priority queue is O($n^{\prime\prime}_p$ $\log(n^{\prime\prime}_p)$). The number of total paths reduces when we apply Pruning Rules~\ref{lemma:pruning2},~\ref{lemma:pruning4},~\ref{lemma:pruning3_1_1},~\ref{lemma:pruning3_1_2}  and~\ref{lemma:pruning5}. Let $r^{\prime\prime}$ be the combined effect factor of these pruning rules. By applying the pruning rules the worst case time complexity of the INE based search becomes $O(\frac{n^{\prime\prime}_p}{r^{\prime\prime}}\log(\frac{n^{\prime\prime}_p}{r^{\prime\prime}}))$.
Thus, the worst case time complexity for the SNVD based approach is $O(k+m + E_q\log(V_q) +\frac{n^{\prime\prime}_p}{r^{\prime\prime}}\log(\frac{n^{\prime\prime}_p}{r^{\prime\prime}}))$.
It can be shown that the cost of finding the $m$ safest paths $O(\frac{n^{\prime\prime}_p}{r^{\prime\prime}}\log(\frac{n^{\prime\prime}_p}{r^{\prime\prime}}))$ dominates the cost of Dijksra's search $O(E_q\log{(V_q)})$ because the search space to find the $m$ safest paths is larger than $VC_q$. Thus, the worst case complexity is 
$O(k+m+\frac{n^{\prime\prime}_p}{r^{\prime\prime}}\log(\frac{n^{\prime\prime}_p}{r^{\prime\prime}}))$
 which is significantly better than the worst case time complexity of INE based approach because $k+m$ is quite small in practice and $n^{\prime\prime}_p/r^{\prime\prime}$ is significantly smaller than  $n_p/r$.}
\subsection {SNVD Update}{\label{SNVD_update}}
An SNVD needs to be updated when there is any change in the edges or POIs of the road network. These changes only have effect on their local neighborhood and thus, the SNVD does not need to be reconstructed from  scratch.  

\emph{\textbf{Adding/removing an edge or a change in the edge safety score:}} When an edge is added or removed or if the ESS changes, we first identify the Voronoi cell $VC$ where the edge is located. Then for every border vertex of $VC$, we recompute the unconstrained safest path between the border vertex and the POI of the Voronoi cell. Any change in the PSS of an unconstrained safest path may require shifting the border vertex location. If the location of a border vertex changes then the adjacent Voronoi cell that also shares the border vertex gets affected. After reconstructing the affected Voronoi cells, we also update their stored information (e.g., the minimum border distance for a border vertex may change for removing an edge).

\emph{\textbf{Adding/removing a POI:}} When a POI is added or removed, we need to first identify the Voronoi cell $VC_x$, where the POI is located. If the POI is removed, the subgraph represented by $VC_x$ is merged with its adjacent Voronoi cells. On the other hand, if a new POI is added then $VC_x$ is divided into two new Voronoi cells and their adjacent Voronoi cells may also get affected. In both cases, we reconstruct the affected Voronoi cells and update their stored information. 

\fi

%% file: sections/7_Experiments.tex
\section{Experiments}{\label{experiments}}

\ifTKDE
\subsection{Setup}
\fi
We propose novel algorithms, $Ct$-tree-$k$SNN and SNVD-$k$SNN 
using a $Ct$-tree and an SNVD, respectively for finding the safest nearby POIs. There is no existing work to compute $k$SNNs and thus, in this paper, we compare $Ct$-tree-$k$SNN and SNVD-$k$SNN 
against two baselines: $R$-tree-$k$SNN and INE-$k$SNN. \revised{$R$-tree-$k$SNN, indexes POIs using an $R$-tree and uses techniques in~\cite{HjaltasonS99} to find the candidate $k$SNN POIs whose Euclidean distances from the query location are less than $d_c$. Then, $R$-tree-$k$SNN applies techniques in~\cite{Islam21} to find the safest valid paths to these candidate POIs to determine the $k$SNNs.} INE-$k$SNN is the extension of INE for evaluating $k$SNN queries (Section~\ref{INE}).



\begin{figure*}[t!]
    \begin{center}
    \begin{tabular}{ccc}
        \hspace{-4mm}
          \resizebox{35mm}{!}{\includegraphics{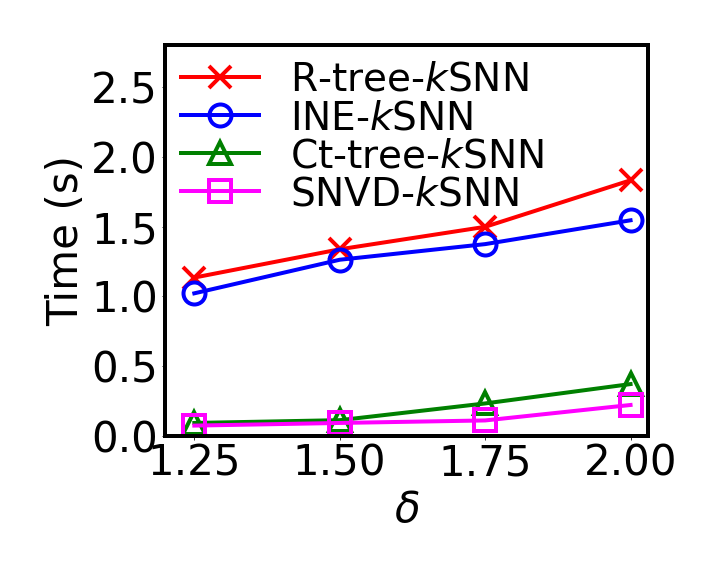}} &
        \hspace{-6mm}
            \resizebox{35mm}{!}{\includegraphics{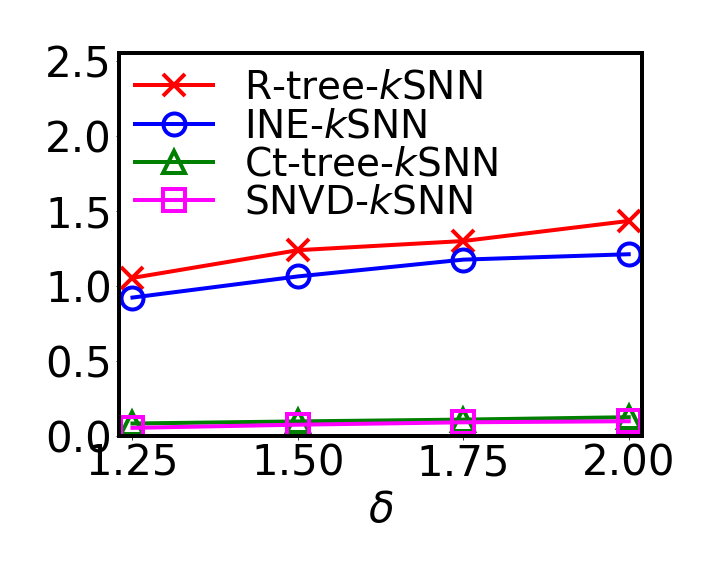}} &
        \hspace{-6mm}
            \resizebox{35mm}{!}{\includegraphics{ 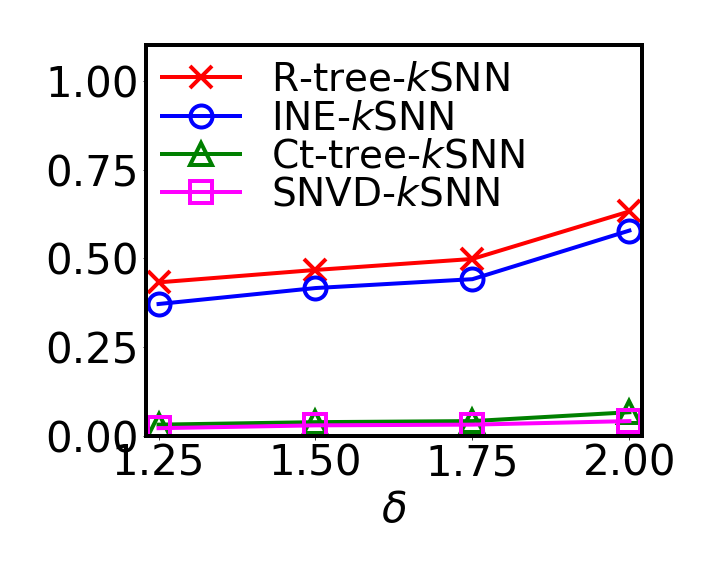}}
         \vspace{-4mm}   
        \\
        \hspace{-4mm}
        \scriptsize{(a) CH \textsc{}\hspace{0mm}} &  \scriptsize{(b) PHL \textsc{}} &  \scriptsize{(c) SF \textsc{}\hspace{0mm}}\\
        
        \hspace{-4mm}
        \resizebox{35mm}{!}{\includegraphics{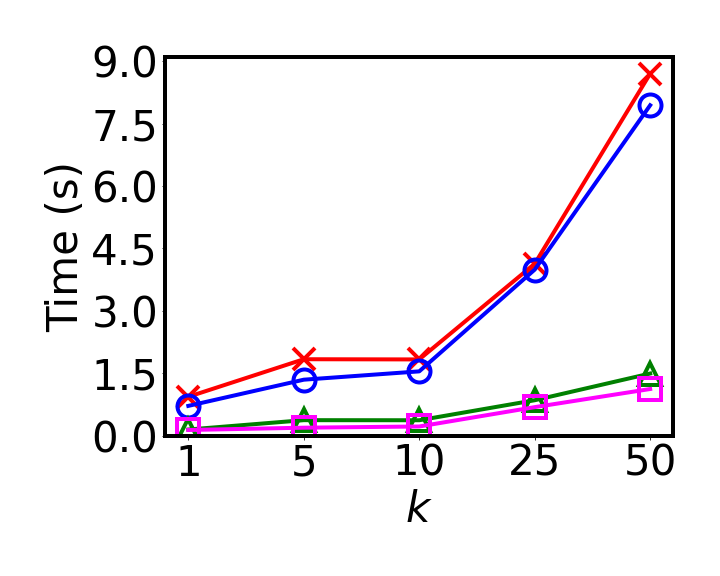}} &
        \hspace{-6mm}
            \resizebox{35mm}{!}{\includegraphics{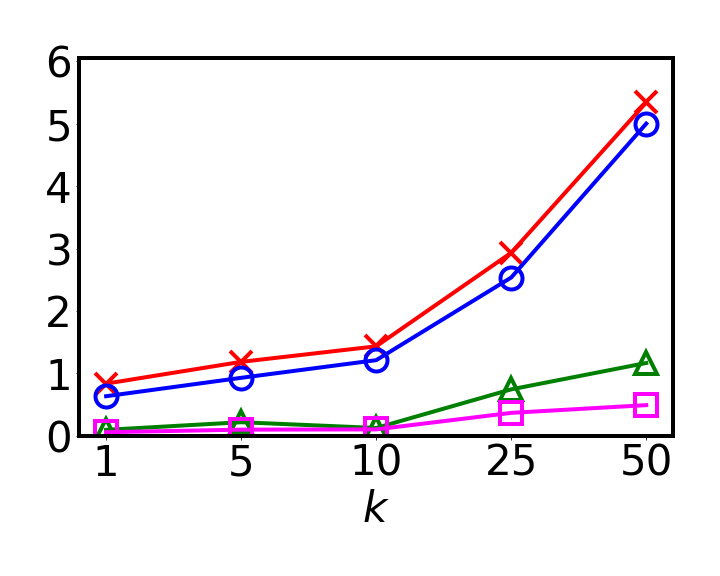}} &
        \hspace{-6mm}
        \resizebox{35mm}{!}{\includegraphics{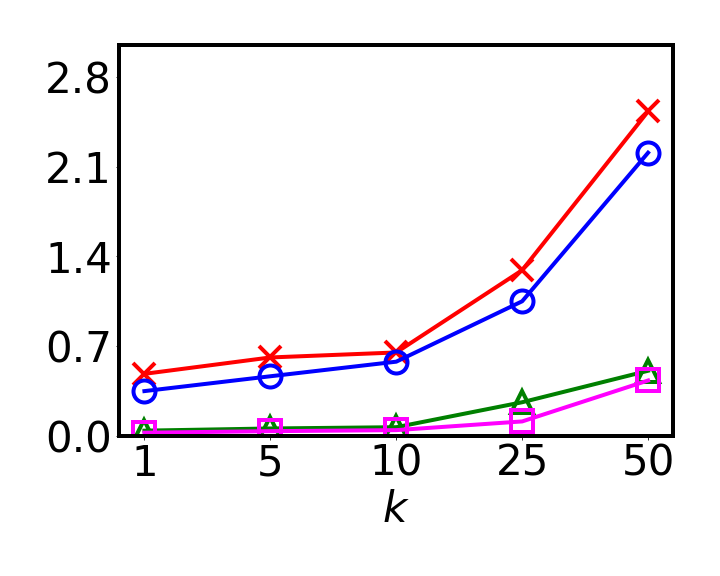}} 
        \vspace{-2mm} 
        \\
        \hspace{-4mm}
        \scriptsize{(d) CH \textsc{}\hspace{0mm}} &  \scriptsize{(e) PHL \textsc{}} &  \scriptsize{(f) SF \textsc{}\hspace{0mm}}\\
     
        \hspace{-2mm}
            \resizebox{35mm}{!}{\includegraphics{ 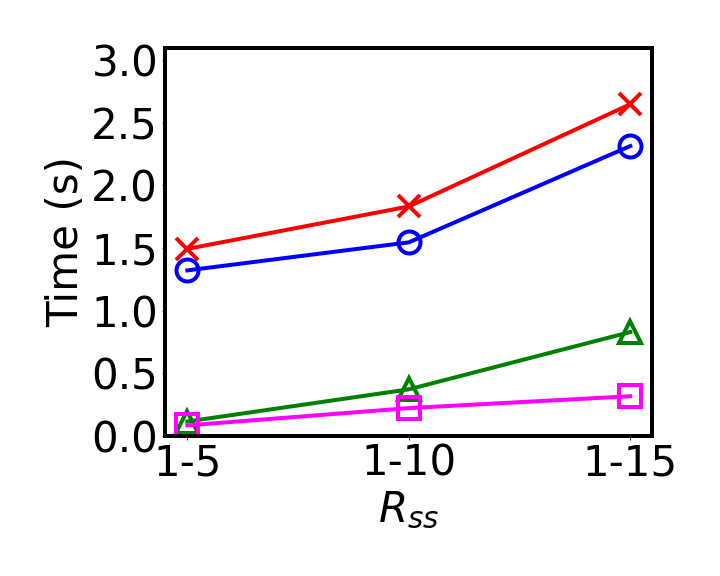}} &
        \hspace{-6mm}
            \resizebox{35mm}{!}{\includegraphics{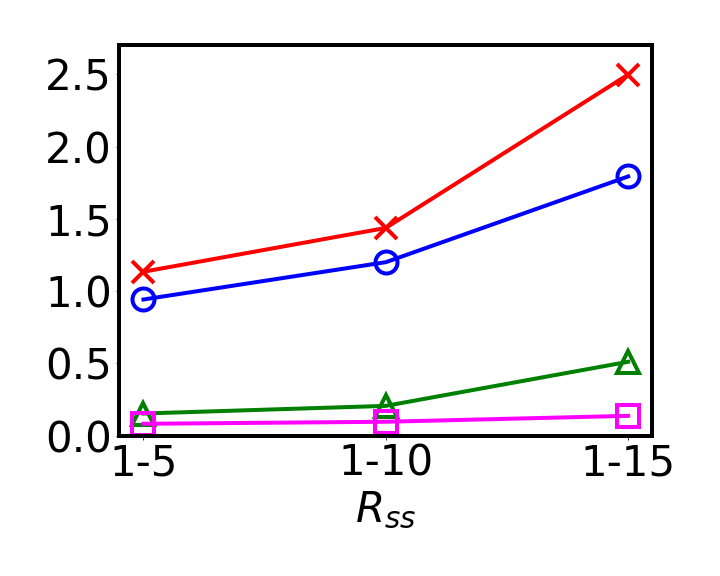}} &
        \hspace{-6mm}
            \resizebox{35mm}{!}{\includegraphics{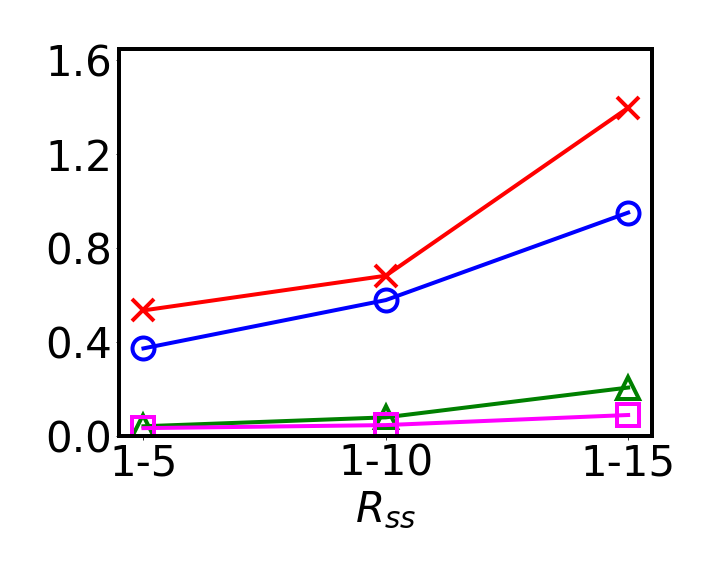}}
        \vspace{-3mm}   
        \\
        \hspace{-1mm}

      \scriptsize{(g) CH \textsc{}\hspace{0mm}} &  \scriptsize{(h) PHL \textsc{}} &  \scriptsize{(i) SF \textsc{}\hspace{0mm}}\\
      
       \vspace{-6mm}
        \hspace{-4mm}
            \resizebox{35mm}{!}{\includegraphics{ 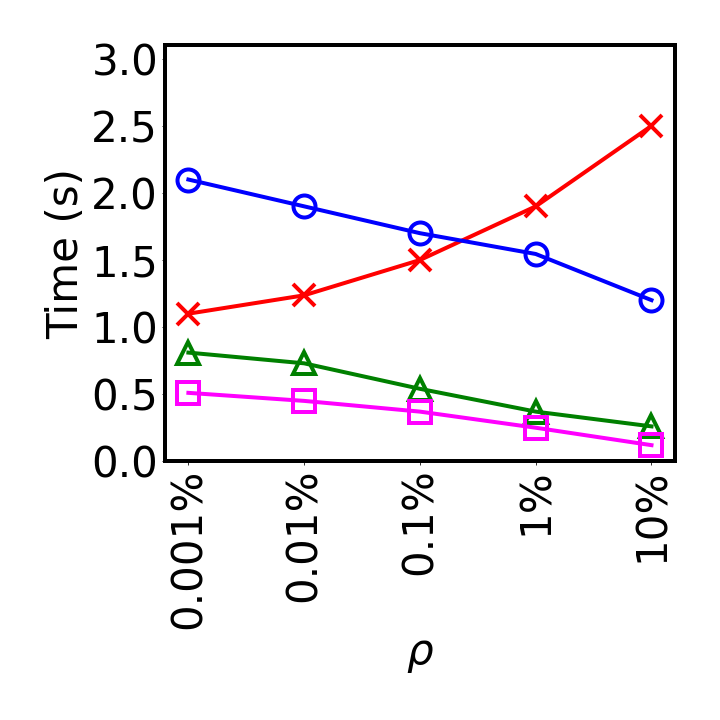}} &
        \hspace{-6mm}
            \resizebox{35mm}{!}{\includegraphics{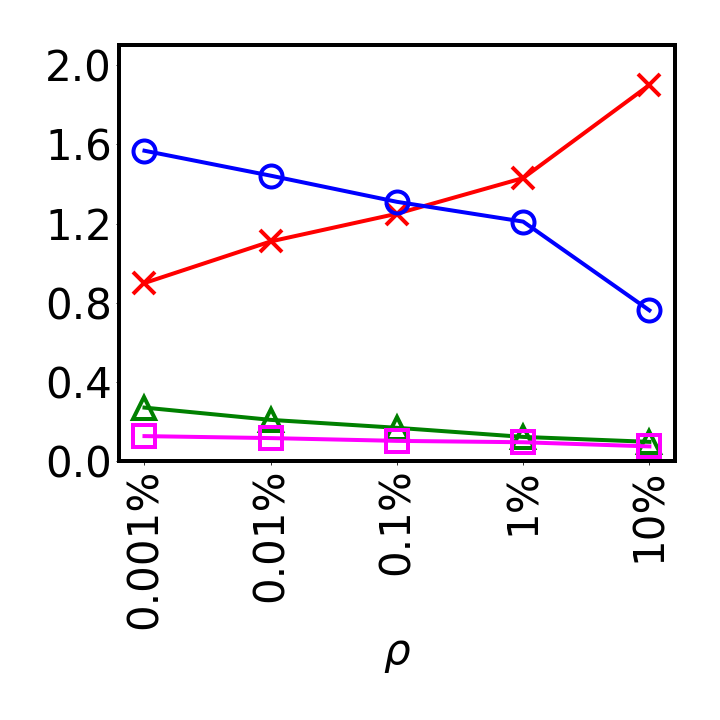}} &
        \hspace{-6mm}
            \resizebox{35mm}{!}{\includegraphics{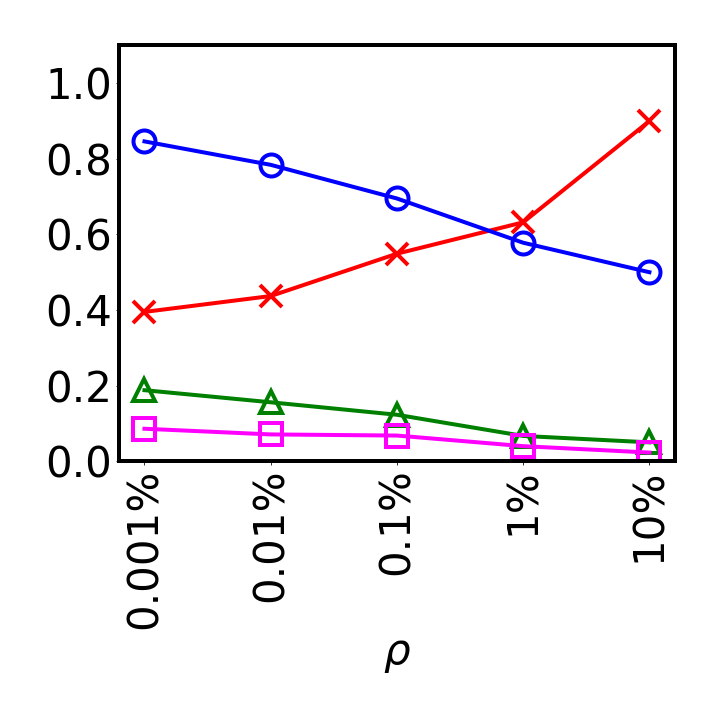}} 
        \vspace{3mm}    
        \\ 
        \hspace{-1mm}

      \scriptsize{(j) CH \textsc{}\hspace{0mm}} &  \scriptsize{(k) PHL \textsc{}} &  \scriptsize{(l) SF \textsc{}\hspace{0mm}}\\

    \end{tabular}
    \caption{\revised{Effect of varying $d_c$, $k$, $R_{ss}$ and $\rho$ for the three datasets}}
    \label{fig:varyParams}
  \end{center}
\end{figure*}

\textbf{\emph{Datasets:}} 
We used real-world datasets of three cities that differ in terms of the road network size and crime statistics. Specifically, we extract the road network of Chicago, Philadelphia and San Francisco, using OpenStreetMap
\ifTKDE
(Table~\ref{Table:Dataset}).
\else
(OSM)\footnote{\url{https://www.openstreetmap.org}} and osm4routing parser\footnote{\url{https://github.com/Tristramg/osm4routing}} in graph format (Table~\ref{Table:Dataset}).
\fi
 All edge weights representing distances were scaled up and integer values were taken. 

\emph{Edge $w_{ij}^{ss}$ Computation.} We used crime datasets of Chicago\footnote{\url{https://data.cityofchicago.org/Public-Safety/Crimes-2001-to-Present/ijzp-q8t2}} \revised{(2001-2019)}, Philadelphia\footnote{\url{https://www.opendataphilly.org/dataset/crime-incidents}} \revised{(2006-2019)} and San Francisco\footnote{\url{https://data.sfgov.org/Public-Safety/Police-Department-Incident-Reports-Historical-2003/tmnf-yvry}} \revised{(2003-2018)} to assign weights ($w_{ij}^{ss}$) representing realistic \revised{ESS}s of the respective road network graphs. \revised{
We generated the ESSs of edges separately for three different ESS ranges $[1,s^{max}]$ denoted as $R_{ss}$ (see Table~\ref{Table:parameters}). 
The crime datasets include locations (latitude and longitude) of the crime incidents like robbery, motor vehicle theft, sexual offense, weapons violation, burglary and criminal damage to vehicle. 
For each edge, we count the number of crime incidents within $1$km and set it as the crime count of the edge. These crime counts are normalized to integers between the range $[1,s^{max}]$ representing crime scores. The crime score of each edge is converted to a safety score as $s^{max}+1-crimescore$, e.g., if $s^{max}$ is $10$ and the crime score for an edge is $10$, the safety score of the edge is $1$. The PSS of a path is computed based on the ESSs included in the path (Definition~\ref{def1}) and can be a float value.}

    \begin{table}
        \centering
         \caption{Road Network Dataset}
            \begin{tabular}{|c|c|c|c|}
                \hline
                \textbf{Dataset}          & \textbf{\# Vertices}      & \textbf{\# Edges}     \\ \hline
                Chicago (\textbf{CH})              &  125,344                &   200,110              \\ \hline
                Philadelphia (\textbf{PHL})        &    80,558               &  120, 581              \\ \hline
                San Francisco (\textbf{SF})        &    40,528                &      72,819               \\ \hline
            \end{tabular} 
            \label{Table:Dataset}
    \end{table}

\textbf{\emph{Parameters:}} 
\revised{The parameter $d_c$ is a user specified parameter. Our solution works for any value of $d_c$. However, in experiments, we set $d_c$ to a value that is large enough to find the $k$SNNs for each query. Specifically, we define $d_c$ to be $\delta\times d^k$ where $d^k$ is the distance between $v_l$ and its $k^{th}$ closest POI and $\delta>1$ is a parameter in the experiments. Existing studies involving other parameterized queries  also adapt similar settings to ensure query results are not empty~\cite{ChenCJW13} and contain at least $k$ POIs~\cite{shao2020efficiently}.} 

We evaluate the algorithms by varying $\delta$, ESS range ($R_{ss}$), $k$, and the density of POIs ($\rho$), where density is the total number of POIs divided by the total number of vertices in the graph. The range and default values of the parameters are shown in Table~\ref{Table:parameters}. Similar to existing works~\cite{ZhongLTZG15,abeywickrama2016k}, \revised{we vary $k$ from 1 to 50} and set default value as 10. 
$R_{ss}$ is varied in the range of [1-5], [1-10] and [1-15]. Note that a smaller range (e.g., [1-2]) cannot effectively distinguish the safety variation of the road segments, whereas for a larger range (e.g., [1-30]), roads with a very small safety variation may have different PSSs and result in an increase in the length of the safest path. As shown in existing work~\cite{abeywickrama2016k}, most of the real world POIs have  pretty low density. Therefore, we evaluate the algorithms by varying $\rho$ from $0.001\%$ to $10\%$ Following~\cite{ZhongLTZG15,abeywickrama2016k}, we set the default value of $\rho$ as 1\%.

    \begin{table}
        \caption{Experiment Parameters}
        \centering
        \resizebox{\linewidth}{!}{
            \begin{tabular}{|c|c|c|}
            \hline
            \textbf{Parameters}                     & \textbf{Range}                            & \textbf{Default}   \\ \hline
                       $k$                             & 1, 5, 10, 25, 50                          & 10                  \\ \hline
            $\delta$                                   & 1.25, 1.5, 1.75, 2                        & 2                \\ \hline
            $R_{ss}$                                   & 1-5, 1-10, 1-15                           & 1-10               \\ \hline
                 $\rho$                                & 0.001\%, 0.01\%, 0.1\%, 1\%, 10\%         & 1\%              \\ \hline
            \end{tabular}}
            \label{Table:parameters}
    \end{table}

\textbf{\emph{Measures:}} For each experiment, we randomly select 100 vertices as query locations and report the average processing time to evaluate a $k$SNN query. We measure the preprocessing and storage cost for $Ct$-tree and SNVD in terms of the construction time, index size and the percentage of border vertices. All algorithms are implemented in C++ and the experiments are run on a 64-bit Windows 7 machine with an Intel Core i3-2350M, 2.30GHz processor and 4GB RAM.

\ifTKDE
\subsection {Results}{\label{queryEvaluation}}
\else
\subsection {Query Performance}{\label{queryEvaluation}}
\fi

\textbf{\emph{Effect of $\delta$: }}
Figures~\ref{fig:varyParams}(a)--\ref{fig:varyParams}(c)  show that our algorithms can find $k$SNNs in real time. The query processing time slowly increases with the increase of $\delta$ (and consequently $d_c$) because more paths need to be explored for larger $\delta$. \revised{For $Ct$-tree-$k$SNN and SNVD-$k$SNN, we observe on average 11.6 times and 13.8 times faster processing time than that of R-tree-$k$SNN and on average 9.7 times and 12.7 times faster processing time than that of INE-$k$SNN, respectively.}

\ifTKDE
\input{sections/table4Short}
\else
\input{sections/table4Long}
\fi

\textbf{\emph{Effect of $k$: }}
Figures~\ref{fig:varyParams}(d)--\ref{fig:varyParams}(f) show that the time increases with the increase of $k$, which is expected. The increase rate is low for $Ct$-tree-$k$SNN and SNVD-$k$SNN and shows almost a linear growth for varying $k$ from 1 to 50,  whereas in case of INE-$k$SNN and \revised{R-tree-$k$SNN}, the time increases significantly. 

\textbf{\emph{Effect of $R_{ss}$: }} Figures~\ref{fig:varyParams}(g)--\ref{fig:varyParams}(i) show that the time increases for \revised{R-tree-$k$SNN}, INE-$k$SNN and $Ct$-tree-$k$SNN for larger $R_{ss}$, whereas for SNVD-$k$SNN the required time is the least among four algorithms and remains almost constant for all $R_{ss}$s. A larger $R_{ss}$ increases the height of the $Ct$-tree which in turn increases the number of subgraphs accessed by $Ct$-tree-$k$SNN. On the other hand, the number of subgraphs represented by SNVD cells does not vary with different $R_{ss}$s.

\textbf{\emph{Effect of $\rho$: }} Figures~\ref{fig:varyParams}(j)--\ref{fig:varyParams}(l) show that the time decreases for INE-$k$SNN, Ct-$k$SNN and SNVD-$k$SNN with the increase of the POI density, $\rho$, \revised{whereas the time sharply increases for R-tree-$k$SNN}. For INE-$k$SNN, a larger $\rho$ increases the possibility of finding the required POIs with less amount of incremental network expansion. For $Ct$-tree-$k$SNN, a larger $\rho$ increases the number of POIs in a $Ct$-tree node which in turn decreases the need of search in a larger subgraph. For SNVD-$k$SNN, the decrease curve is less steeper, since in SNVD-$k$SNN, the search time mainly depends on the number of SNVD cells that need to be explored rather than the total number of SNVD cells in the SNVD.  \revised{On the other hand, for R-tree-$k$SNN, the number of candidates for $k$SNNs increases with the increase in $\rho$, which also causes increase in the number of independent safest path computations.}

\textbf{\emph{Effect of Datasets and Performance Scalability: }}
In our experiments, we used three datasets that vary in the number of road network vertices, edges, POIs and crime distributions. Irrespective of the datasets, both $Ct$-tree-$k$SNN and SNVD-$k$SNN outperform \revised{R-tree-$k$SNN and} INE-$k$SNN  with a large margin. In addition, the query processing time linearly increases with increase of the dataset size in terms of the number of road network vertices, edges and POIs. Thus, our algorithms are scalable and applicable for all settings. In default parameter setting, the average time for Ct-kSNN is 371 ms, 123 ms and 66 ms for CH, PHL, and SF datasets, respectively. The average time for SNVD-kSNN is 223 ms, 96 ms and 41 ms for for CH, PHL, and SF datasets, respectively.


\textbf{\emph{Pruning Effectiveness:}} 
In this set of experiments, we investigate the effectiveness of our pruning rules for the search space refinement. \revised{Table~\ref{Table:Pruning}
summarizes the experiment results in the default setting of the parameters in terms of the query processing time, number of network vertices accessed $\|V_a\|$ and number of valid paths explored in road networks for different approaches when different pruning rules are applied. The Basic  Algorithm includes only Pruning Rule~\ref{lemma:pruning1}(i.e., considers only valid paths).} The performance of our algorithms are the best when all pruning rules are applied, which in turn means that each of the pruning rule contributes in improving the efficiency of our algorithms.
\ifTKDE
\revised{Table~\ref{Table:Pruning}  also explains why SNVD is faster than $Ct$-tree and $Ct$-tree is faster than INE based approach.}
\else
\revised{Table~\ref{Table:Pruning} also shows why SNVD outperforms the other two algorithms. Note that the number of valid paths explored by the Basic Algorithm corresponds to $n_p$, $n^\prime_p$ and $n^{\prime\prime}_p$ in the complexity analyses of INE, $Ct$-tree and SNVD, respectively. Applying all pruning rules reduces the number of paths explored. The percentage values w.r.t. the Basic Algorithm are shown in parentheses. These percentage values can be used to obtain effect factors $r$, $r^\prime$ and $r^{\prime\prime}$ in the complexity analyses (e.g., $50\%$ reduction in number of paths implies that the effect factor is $100/50=2$). Generally, $n^{\prime\prime}_p > n^\prime_p > n_p$ and $r^{\prime\prime} > r^{\prime} > r$ which explains why SNVD is faster than $Ct$-tree and $Ct$-tree is faster than INE based approach. }

\ifTKDE
\input{sections/table5Short}
\else
\input{sections/table5Long}
\fi 

\ifTKDE
\revised{
\textbf{\emph{Query Effectiveness:}}  }
\else
\revised{
\subsection{Query Effectiveness}
}
\fi
\revised{
\ifTKDE
We compare $k$SNNs with traditional $k$ nearest neighbors ($k$NNs) on CH dataset using default settings but varying $\delta$. The results on the other datasets can be found in~\cite{biswas2021safest}. We generate 100 query locations and run a kNN query and a kSNN query for each of these query locations, and report average results. 
\else
We compare $k$SNNs with traditional $k$ nearest neighbors ($k$NNs) using default settings but varying $\delta$. Specifically, for each data set, we generate 100 query locations and run a kNN query and a kSNN query for each of these query locations, and report average results.
\fi
We compare $k$SNNs with traditional $k$ nearest neighbors ($k$NNs) using default settings but varying $\delta$. Specifically, for each data set, we generate 100 query locations and run a kNN query and a kSNN query for each of these query locations, and report average results. Table~\ref{Table:Effectiveness} shows that $k$SNNs are significantly different from $k$NNs (see number of common POIs). Furthermore, the traditional $k$NN queries are not likely to include a POI that is also the first SNN. Finally, although the path lengths to $k$SNNs are longer, these paths are much safer.
\ifTKDE
Specifically, the paths to $k$SNNs are up to $1.6$ times longer on average but they are up to $16$ times safer than the paths to $k$NNs.
\else
Specifically, the paths to $k$SNNs are up to $1.76$ times longer on average but they are up to $21$ times safer than the paths to $k$NNs.
\fi
We also observe that the number of common POIs among $k$SNNs and $k$NNs decreases and the safety (PSS) of the paths from $v_l$ to $k$SNNs increases with the increase of $\delta$. Thus, in real-world scenarios where safety may be important for users, an application may show some $k$SNNs as well as some $k$NNs to give users more options to choose from.}

\ifTKDE

\revised{
\textbf{\emph{Preprocessing and Index Size:}}  
We also evaluated the construction time and index size of $Ct$-tree and SNVD. The detailed results can be found in~\cite{biswas2021safest}. We observe that the construction times for $Ct$-tree and SNVD are comparable whereas SNVD is up to 1.6 times bigger in size than $Ct$-tree. Specifically, for default settings (i.e., $R_{ss}=1-10$, $\rho=1\%$), the construction time for $Ct$-tree for the three datasets ranges from $54-123$ minutes as compared to $50-122$ minutes for SNVD. On the other hand, the index size of $Ct$-tree for the three datasets ranges from $63-114$ MB compared to $101-168$ MB for SNVD.}

\else
\subsection{Preprocessing and Storage}
Query parameters $d_c$ and $k$ do not affect preprocessing and storage costs of the $Ct$-tree and  SNVD. We show the effect of  $R_{ss}$, $\rho$ and datasets on the preprocessing and storage costs of the $Ct$-tree and  SNVD in experiments.

\textbf{\emph{$Ct$-tree:}} 
Figure~\ref{fig:ConsAndStorageCt} shows that the $Ct$-tree construction time, index size and the percentage of border vertices slowly increase with the increase of $R_{ss}$ and dataset size. 
 The effect of dataset size on the increasing trend is expected, whereas the reason behind the increase of processing time and storage for $R_{ss}$ is that the height of a $Ct$-tree increases with the increase of $R_{ss}$. The $Ct$-tree stores 1\% to 5\% of the total number of vertices in the road network as border vertices, which is negligible. We observed that the density of POIs $\rho$  does not affect construction cost as the construction of $Ct$-tree does not depend on the number of POIs but mainly on the \revised{ESS} values of the edges.
 While border to POI distances need to be maintained, the effect on index size is not significant. 
 
\begin{small}
\begin{figure}[t]
    \begin{center}
    \begin{tabular}{ccc}
        \hspace{-5mm}
        \resizebox{28mm}{!}{\includegraphics{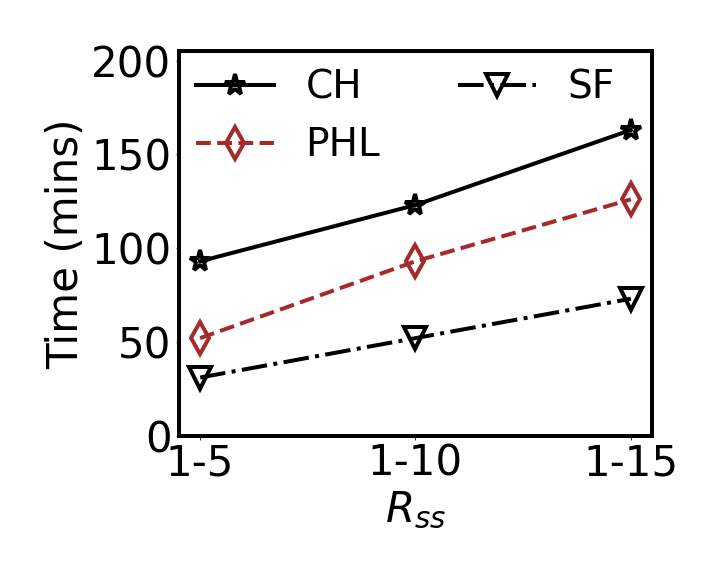}} &
        \hspace{-6mm}
        \resizebox{28mm}{!}{\includegraphics{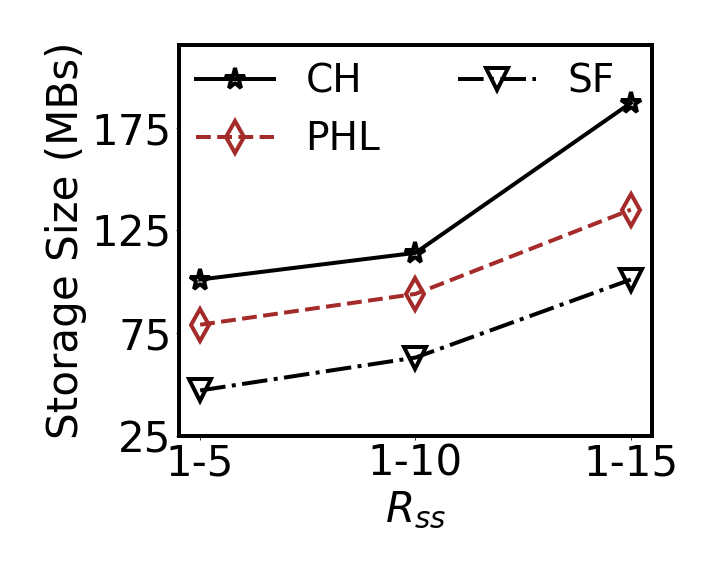}} &
        \hspace{-6mm}  
        \resizebox{28mm}{!}{\includegraphics{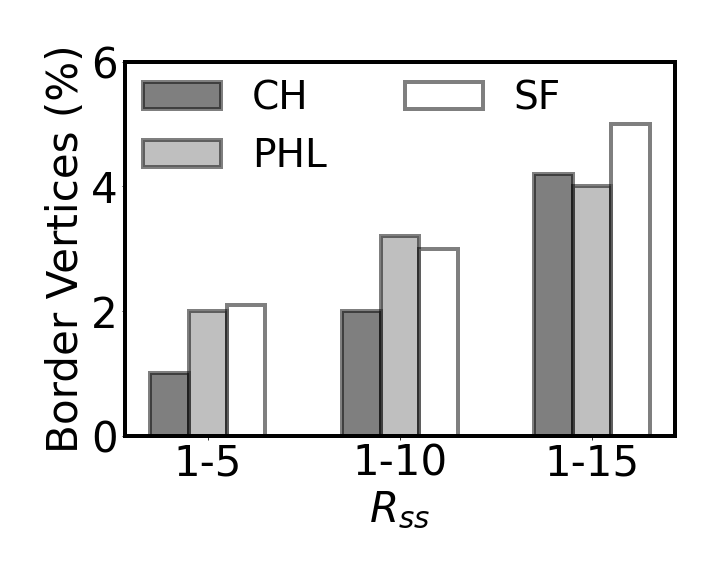}\vspace{-3mm}}
        \vspace{-2mm}
        \\
        \hspace{-1mm}
        \scriptsize{(a) \textsc{}\hspace{0mm}} &  \scriptsize{(b) \textsc{}} &  \scriptsize{(c) \textsc{}\hspace{0mm}}\\
    \end{tabular}
    \vspace{-2mm}
    \caption{\textbf{Ct-tree:} Effect of $R_{ss}$ and datasets on \revised{(a) construction time (b) storage size} and (c) \% of border vertices}
    \label{fig:ConsAndStorageCt}
    \end{center}
\vspace{-4mm}
\end{figure}
\end{small}

\begin{figure}[t]
    \begin{center}
    \begin{tabular}{ccc}
        \hspace{-6mm}
        
        \resizebox{28mm}{!}{\includegraphics{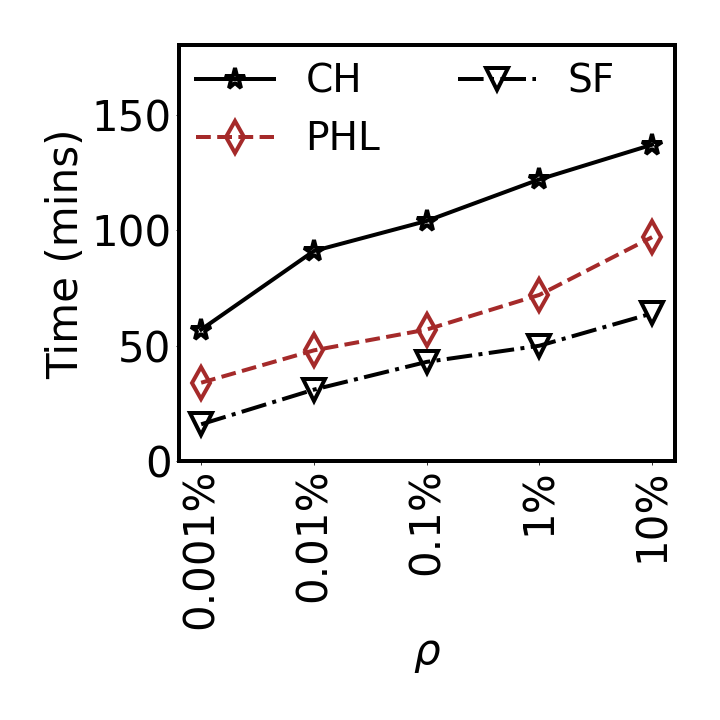}} &
        \hspace{-6mm}
        \resizebox{28mm}{!}{\includegraphics{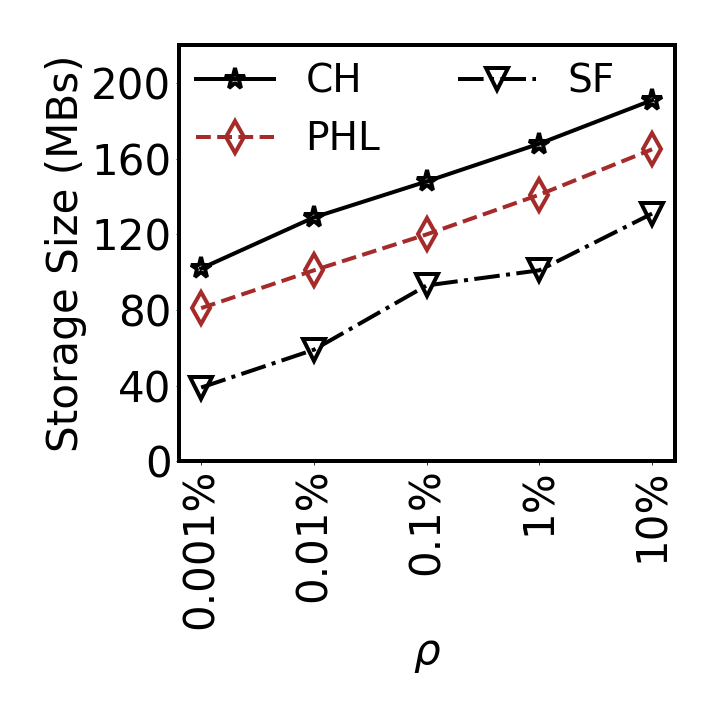}} &
        \hspace{-6mm}  
        \resizebox{28mm}{!}{\includegraphics{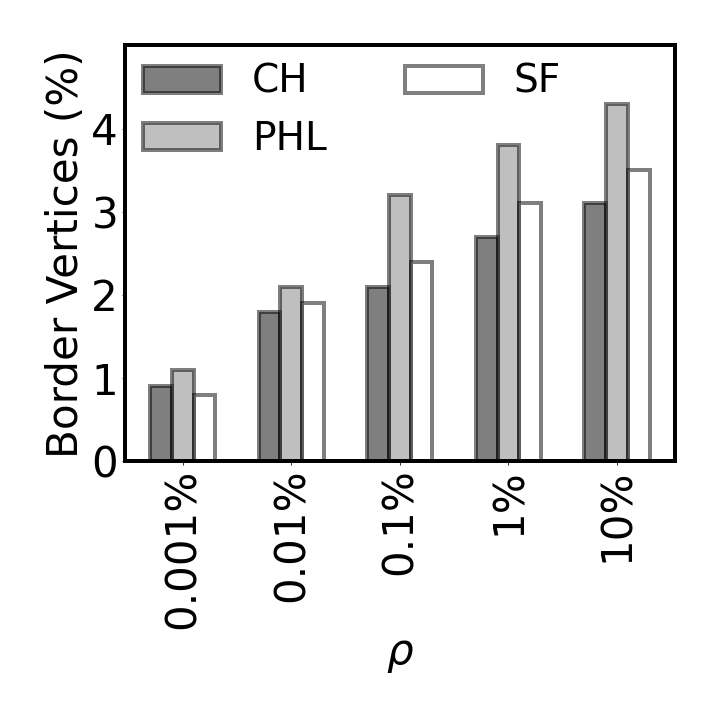}}
        \vspace{-2mm}
        \\
        \hspace{-1mm}
        \scriptsize{(a) \textsc{}\hspace{0mm}} &  \scriptsize{(b) \textsc{}} &  \scriptsize{(c) \textsc{}\hspace{0mm}}\\
    \end{tabular}
    \caption{\textbf{SNVD:} Effect of $\rho$ and datasets on \revised{(a) construction time (b) storage size} and (c) \% of border vertices}
    \label{fig:ConsAndStorageVN}
    \end{center}
\end{figure}

\textbf{\emph{SNVD:}} The storage size of an SNVD includes SNVD cell information \revised{and distance information}. 
We observe no significant variation in the construction time, storage size and the number of border vertices for different \revised{ESS} ranges $R_{ss}$. The number of Voronoi cells in an SNVD is equal to the number of POIs and thus, the construction time, storage size and the number of border vertices increase with the increase of $\rho$ and the dataset size (Figure~\ref{fig:ConsAndStorageVN}). The percentage of border vertices for all datasets is less than 5\% of the total number of vertices in the road network, which is negligible. 

\revised{
\textbf{\emph{$Ct$-tree vs. SNVD:}}  
The construction times for $Ct$-tree and SNVD are comparable whereas SNVD is up to 1.6 times bigger in size than $Ct$-tree. Specifically, for default settings (i.e., $R_{ss}=1-10$, $\rho=1\%$), the construction time for $Ct$-tree for the three datasets ranges from $54-123$ minutes as compared to $50-122$ minutes for SNVD. On the other hand, the index size of $Ct$-tree for the three datasets ranges from $63-114$ MB compared to $101-168$ MB for SNVD.}

\fi

%% file: sections/table4Short.tex
\begin{small} 
\begin{table}[htbp]
\centering
\caption{\revised{Effectiveness of Pruning Rules (PRules) in terms of the query processing time, \# of vertices accessed $|V_a|$ and \# of valid paths explored by the algorithms on CH dataset. The Basic algorithm in each approach only applies Pruning Rule~\ref{lemma:pruning1}. Each \% value $x$ shown in a parenthesis indicates that the value in this cell is $x\%$ of the value of the Basic of the same approach.}}
\resizebox{1\columnwidth}{!}{
{
\ifTKDE
\color{black}
\else 
\color{black}
\fi
\begin{tabular}{|c|c|c|c|c|}
\hline
 & \textbf{Setting} & \textbf{Time (ms)} & \textbf{$|V_a|$} & \textbf{\begin{tabular}[c]{@{}c@{}}\# of Valid\\  Paths\end{tabular}}  \\ \hline \hline
       & Basic  & 2225               & 2708            & 18223                                                                 \\ \cline{2-5}
 \multirow{-2}{*}{INE}     &  PRule~\ref{lemma:pruning2}       & 1787 (80.31\%)        & 2545 (93.9\%)     & 14534 (79.75\%) \\ \hline \hline           
    & Basic  & 682                & 1601            & 4531                                                  \\ \cline{2-5} 
  &  PRule~\ref{lemma:pruning2}       & 485 (71.11\%)            & 1056 (65.95\%)     & 3287 (72.65\%)                                                  \\ \cline{2-5} 
   & PRule~\ref{lemma:pruning3}     & 460 (67.44\%)          & 942 (58.83\%)      & 2721 (60.05\%)                                                          \\ \cline{2-5} 
   & PRule~\ref{lemma:pruning4}  & 471 (69.06\%)          & 955 (59.65\%)      & 2913 (64.31 \% )      \\ \cline{2-5} 
\multirow{-5}{*}{$Ct$-tree} & All      & 377 (55.27\%)          & 875 (54.65 \% )      & 2621 (57.84\%)\\ \hline \hline
   & Basic  & 442                & 391             & 942                                                      \\ \cline{2-5} 
  &  PRule~\ref{lemma:pruning2}       & 349 (78.95\%)         & 341 (87.21\%)      & 699 (74.20 \% )    \\ \cline{2-5} 
  & PRules~\ref{lemma:pruning3_1_1},~\ref{lemma:pruning3_1_2}  & 364 (82.35\%)          & 356 (91.04\%)      & 719 (76.32 \% ) \\ \cline{2-5} 
   & PRule~\ref{lemma:pruning5}   & 379  (85.74\%)          & 371 (94.88\%)      & 772 (81.95\%)       \\ \cline{2-5} 
\multirow{-5}{*}{SNVD}    & All      & 229 (51.81\%)        & 162 (58.56\%)      & 623 (66.13\%)\\ \hline
\end{tabular}
}
\label{Table:Pruning}

\end{table}
\end{small}

%% file: sections/table4Long.tex
\begin{table*}[htbp]
\centering
\caption{\revised{Pruning effectiveness in terms of the query processing time, number of network vertices accessed $\mid V_a\mid$ and number of valid paths explored by the algorithms. The Basic Algorithm in each approach only considers valid paths (i.e., only applies Pruning Rule~\ref{lemma:pruning1}). Each $\%$ value $x$ shown in a parenthesis indicates that the value in this cell is $x\%$ of the value of the Basic Algorithm of the same approach.}}

\resizebox{1\textwidth}{!}{
{
\ifTKDE
\color{blue}
\else 
\color{black}
\fi

\begin{tabular}{cc|c|c|c|c|c|c|c|c|c|}
\cline{3-11}
     &                  & \multicolumn{3}{c|}{\textbf{CH}}                                                                             & \multicolumn{3}{c|}{\textbf{PHL}}                                                                            & \multicolumn{3}{c|}{\textbf{SF}}                                                                             \\ \hline
\multicolumn{1}{|c|}{\textbf{}}             & \textbf{Setting} & \textbf{Time (ms)} & \textbf{$\mid V_a \mid$} & \textbf{\begin{tabular}[c]{@{}c@{}}\# of Valid\\  Paths\end{tabular}} & \textbf{Time (ms)} & \textbf{$\mid V_a \mid $} & \textbf{\begin{tabular}[c]{@{}c@{}}\# of Valid\\  Paths\end{tabular}} & \textbf{Time (ms)} & \textbf{$\mid V_a \mid $} & \textbf{\begin{tabular}[c]{@{}c@{}}\# of Valid\\  Paths\end{tabular}} \\ \hline \hline
\multicolumn{1}{|c|}{}                               & Basic Algorithm  & 2225               & 2708            & 18223                                                                 & 755               & 2438            & 15867                                                                 & 1140               & 1905            & 12232                                                                 \\ \cline{2-11} 
\multicolumn{1}{|c|}{\multirow{-2}{*}{INE}}     &  Pruning Rule~\ref{lemma:pruning2}       & 1787 (80.31\%)        & 2545 (93.9\%)     & 14534 (79.75\%)                                                          & 581 (76.95\%)        & 1831 (75.1\%)      & 12783 (80.71\%)                                                          & 887 (77.8\%)          & 1701 (77.5\%)      & 8536 (69.78 \% )                                                           \\ \hline \hline
\multicolumn{1}{|c|}{}                               & Basic Algorithm  & 682                & 1601            & 4531                                                                  & 204                & 732             & 3673                                                                  & 109                & 351             & 2321                                                                  \\ \cline{2-11} 
\multicolumn{1}{|c|}{}                               &  Pruning Rule~\ref{lemma:pruning2}       & 485 (71.11\%)            & 1056 (65.95\%)     & 3287 (72.65\%)                                                          & 192 (94.11\%)        & 601 (82.1\%)      & 3328 (90.62\%)                                                         & 84 (82.35\% )       & 318 (90.59\%)     & 1881 (81.04\%)                                                          \\ \cline{2-11} 
\multicolumn{1}{|c|}{}                               & Pruning Rule~\ref{lemma:pruning3}     & 460 (67.44\%)          & 942 (58.83\%)      & 2721 (60.05\%)                                                          & 194 (95.09\%)       & 606 (82.7\% )      & 3451 (93.97\%)                                                          & 81 (79.41 \% )        & 302 (86.03\%)     & 1745 (75.18\% )                                                          \\ \cline{2-11} 
\multicolumn{1}{|c|}{}                               & Pruning Rule~\ref{lemma:pruning4}  & 471 (69.06\%)          & 955 (59.65\%)      & 2913 (64.31 \% )                                                          & 188 (92.15\%)        & 588 (80.32\%)     & 3198 (87.06\%)                                                          & 86 (81.13\%)        & 315 (89.74\%)     & 1801 (77.59\%)                                                          \\ \cline{2-11} 
\multicolumn{1}{|c|}{\multirow{-5}{*}{$Ct$-tree}} & All Pruning      & 377 (55.27\%)          & 875 (54.65 \% )      & 2621 (57.84\%)                                                          & 131 (64.8\%)        & 523 (71.72\%)     & 2665 (72.48\%)                                                          & 70 (68.62\%)        & 281 (80.05\%)     & 1535 (66.13\%)                                                          \\ \hline \hline
\multicolumn{1}{|c|}{}                               & Basic Algorithm  & 442                & 391             & 942                                                                   & 145                & 281             & 625                                                                   & 102                & 194             & 732                                                                   \\ \cline{2-11} 
\multicolumn{1}{|c|}{}                               &  Pruning Rule~\ref{lemma:pruning2}       & 349 (78.95\%)         & 341 (87.21\%)      & 699 (74.20\% )                                                           & 120 (82.75\%)       & 249 (88.61\%)    & 538 (86.08\%)                                                           & 78 (75.68\%)         & 150 (77.31\%)     & 627 (85.65\%)                                                           \\ \cline{2-11} 
\multicolumn{1}{|c|}{}                               & Pruning Rules~\ref{lemma:pruning3_1_1},~\ref{lemma:pruning3_1_2}  & 364 (82.35\%)          & 356 (91.04\%)      & 719 (76.32 \% )                                                         & 126 (84.13\%)        & 255 (90.74\%)     & 569 (91.04\%)                                                           & 81 (79.28\%)         & 155 (79.89\%)     & 639 (87.29\%)                                                           \\ \cline{2-11} 
\multicolumn{1}{|c|}{}                               & Pruning Rule~\ref{lemma:pruning5}   & 379  (85.74\%)          & 371 (94.88\%)      & 772 (81.95\%)                                                           & 122 (84.13\%)        & 205 (72.95\%)    & 485 (77.6\%)                                                            & 71 (69.37\%)      & 151 (77.84\%)     & 643 (87.84\%)                                                           \\ \cline{2-11} 
\multicolumn{1}{|c|}{\multirow{-5}{*}{SNVD}}    & All Pruning      & 229 (51.81\%)        & 162 (58.56\%)      & 623 (66.13\%)                                                          & 96 (66.21\%)         & 165 (58.71\%)     & 401 (64.16\%)                                                           & 53 (51.96\%))        & 93 (47.94\%)     & 409 (55.87\%)                                                           \\ \hline
\end{tabular}
}}
\label{Table:Pruning}

\end{table*}

%% file: sections/table5Short.tex
\begin{table}[t]
\centering
\caption{\revised{For $k=10$, comparison of $k$NN and $k$SNN queries on CH dataset showing: average \# of common POIs in $k$NNs and $k$SNNs; probability that $1$SNN is one of the POIs in $k$NNs; average PSS of the $k$ paths returned by $k$NN queries vs $k$SNN queries, and their ratios (i.e., $k$NN:$k$SNN); and average length of the $k$ paths returned by $k$NN vs $k$SNN queries, and their ratios.} }
\resizebox{1\columnwidth}{!}{
{
\ifTKDE
\color{black}
\else 
\color{black}
\fi
\begin{tabular}{cc|c|c|c|c|}
 \hline
 \multicolumn{2}{|c|}{Measure} & $\delta$=1.25 & $\delta$ =1.5 & $\delta$ = 1.75 & $\delta$ = 2  \\ \hline \hline
 \multicolumn{2}{|c|}{\#common POIs} & 4.71 & 4.12 & 2.23 & 1.34  \\ \hline \hline
 \multicolumn{2}{|c|}{\begin{tabular}[c]{@{}c@{}}probability of \\$1$SNN in $k$NN\end{tabular}} & 0.19 & 0.15 & 0.13 & 0.12  \\ \hline \hline
 \multicolumn{1}{|c|}{\multirow{2}{*}{PSS}} & $k$NN & 0.00029  & 0.00029 & 0.00029 & 0.00029  \\ \cline{2-6}
 \multicolumn{1}{|c|}{\multirow{2}{*}{}} & $k$SNN & 0.00091  & 0.0018 & 0.0024 & 0.0048 \\ \cline{2-6}
\multicolumn{1}{|c|}{\multirow{2}{*}{}} & ratio & 1:3.1  & 1:6.2 & 1:8.3 & 1:16.5  \\ \hline \hline \multicolumn{1}{|c|}{\multirow{2}{*}{Length}} & $k$NN & 1311  & 1311 & 1311 & 1311 \\ \cline{2-6}
\multicolumn{1}{|c|}{\multirow{2}{*}{}} & $k$SNN & 1507  & 1717 & 1941 & 2084 \\ \cline{2-6}
 \multicolumn{1}{|c|}{\multirow{2}{*}{}} & ratio & 1:1.15  & 1:1.3 & 1:1.5 & 1:1.6  \\ \hline 
\end{tabular}
}}
\label{Table:Effectiveness}
\end{table}

%% file: sections/table5Long.tex
\begin{table*}[t]
\centering
\caption{\revised{For $k=10$, comparison of $k$NN and $k$SNN queries including: average \# of common POIs in $k$NNs and $k$SNNs; probability that $1$SNN is one of the POIs in $k$NNs; average PSS of the $k$ paths returned by $k$NN queries vs $k$SNN queries, and their ratios (i.e., $k$NN:$k$SNN); and average length of the $k$ paths returned by $k$NN vs $k$SNN queries, and their ratios.} }
\resizebox{1\textwidth}{!}{
{
\ifTKDE
\color{blue}
\else 
\color{black}
\fi
\begin{tabular}{cc|c|c|c|c|c|c|c|c|c|c|c|c|}
\cline{3-14} & &  \multicolumn{4}{c|}{\textbf{CH}}  & \multicolumn{4}{c|}{\textbf{PHL}} & \multicolumn{4}{c|}{\textbf{SF}} \\ \hline

\multicolumn{2}{|c|}{Measure} & $\delta$=1.25 & $\delta$ =1.5 & $\delta$ = 1.75 & $\delta$ = 2 & $\delta$=1.25 & $\delta$=1.5 & $\delta$ = 1.75 & $\delta$ = 2 & $\delta$=1.25 & $\delta$ =1.5 & $\delta$ = 1.75 & $\delta$ = 2 \\ \hline \hline
\multicolumn{2}{|c|}{\#common POIs} & 4.71 & 4.12 & 2.23 & 1.34 & 5.43 & 4.35 & 4.12 & 2.15 & 4.33 & 3.71 & 3.32 & 1.45 \\ \hline \hline
\multicolumn{2}{|c|}{\begin{tabular}[c]{@{}c@{}}probability of \\$1$SNN in $k$NN\end{tabular}} & 0.19 & 0.15 & 0.13 & 0.12 & 0.21 & 0.18 & 0.14 & 0.10 & 0.18 & 0.15 & 0.13 & 0.11 \\ \hline \hline
\multicolumn{1}{|c|}{\multirow{2}{*}{PSS}} & $k$NN & 0.00029  & 0.00029 & 0.00029 & 0.00029 & 0.000035 & 0.000035 & 0.000035 & 0.000035 & 0.0004 & 0.0004 & 0.0004 & 0.0004 \\ \cline{2-14}
\multicolumn{1}{|c|}{\multirow{2}{*}{}} & $k$SNN & 0.00091  & 0.0018 & 0.0024 & 0.0048 & 0.00006 & 0.00014 & 0.00045 & 0.00076 & 0.00055 & 0.00093 & 0.0012 & 0.003 \\ \cline{2-14}
\multicolumn{1}{|c|}{\multirow{2}{*}{}} & ratio & 1:3.1  & 1:6.2 & 1:8.3 & 1:16.5 & 1:1.7 & 1:4 & 1:12.8 & 1:21.7 & 1:1.4 & 1:2.3 & 1:3 & 1:7.6 \\ \hline \hline
\multicolumn{1}{|c|}{\multirow{2}{*}{Length}} & $k$NN & 1311  & 1311 & 1311 & 1311 & 1222 & 1222 & 1222 & 1222 & 2446 & 2446 & 2446 & 2446 \\ \cline{2-14}
\multicolumn{1}{|c|}{\multirow{2}{*}{}} & $k$SNN & 1507  & 1717 & 1941 & 2084 & 1504 & 1650 & 1882 & 2150 & 2960 & 3302 & 3717 & 3791 \\ \cline{2-14}
\multicolumn{1}{|c|}{\multirow{2}{*}{}} & ratio & 1:1.15  & 1:1.3 & 1:1.5 & 1:1.6 & 1:1.23 & 1:1.35 & 1:1.54 & 1:1.76 & 1:1.21 & 1:1.35 & 1:1.52 & 1:1.55 \\ \hline 
\end{tabular}
}}
\label{Table:Effectiveness}
\end{table*}

%% file: sections/8_Conclusion.tex
\section{Conclusions and Future Work}{\label{conclusion}}
\revised{In this paper, we introduced $k$ Safest Nearby neighbor ($k$SNN) queries in road networks and formulated the measure of path safety score. We proposed novel indexing structures, $Ct$-tree and a safety score based Voronoi diagram (SNVD), to efficiently evaluate $k$SNN queries. We adapt the INE based technique and the $R$-tree based technique to develop two baselines to find $k$SNNs.  Our extensive experimental study on three real-world datasets shows that $Ct$-tree and SNVD based approaches are up to an order of magnitude faster than the baselines.  Comparing $Ct$-tree and SNVD on default settings, both approaches have comparable construction time whereas SNVD index is around 1.6 times bigger than $Ct$-tree but is 1.2-1.7 times faster in terms of query processing cost. Thus, if the storage is not an issue, one should go for the SNVD based approach to get faster query processing performance.
}

\revised{An interesting direction for future work is to investigate skyline routes~\cite{GalbrunPT16,josse2015tourismo} considering multiple criteria such as distance, safety, travel time etc. Here, the safety itself may consist of multiple attributes corresponding to different crime types and inconveniences, e.g., robbery, harassment, bumpy road etc.}

\noindent
\textbf{Acknowledgments.}
Tanzima Hashem is supported by basic research grant of Bangladesh University of Engineering and Technology. Muhammad Aamir Cheema is supported by the Australian Research Council (ARC) FT180100140 and DP180103411.

%% file: sections/9_Appendix.tex
\section{Proof of Lemma~\ref{ss_property2}}
\label{appendix1}
\begin{proof}
From Definition~\ref{def_PSS},  $pss(pt_{lj}) = \frac{1}{\sum_{s=1}^{s^{max}}{w_s \times pt_{lj}.d_s}}$. Replacing $pt_{lj}.d_s$ according to Definition~\ref{def_ds}, we have
\begin{equation*}
  pss(pt_{lj}) = \frac{1}{\sum_{s=1}^{s^{max}}{w_s \times \sum_{e_{xy}\in pt_{lj} \wedge w_{xy}^{ss}=s} w^d_{xy}}}
 \end{equation*}
 
 \begin{equation*}
   = \frac{1}{\sum_{s=1}^{s^{max}}{w_s \times (\sum_{e_{xy}\in pt_{li} \wedge w_{xy}^{ss}=s} w^d_{xy} + \sum_{e_{xy}\in pt_{ij} \wedge w_{xy}^{ss}=s} w^d_{xy}})} 
\end{equation*}
\begin{equation*}
  =\frac{1}{\sum_{s=1}^{s^{max}}{(w_s \times \sum_{e_{xy}\in pt_{li} \wedge w_{xy}^{ss}=s} w^d_{xy}})+(w_s \times \sum_{e_{xy}\in pt_{ij} \wedge w_{xy}^{ss}=s} w^d_{xy})} 
 \end{equation*}
 \begin{equation*}
  =\frac{1}{(\sum_{s=1}^{s^{max}}{w_s \times pt_{li}.d_s})+(\sum_{s=1}^{s^{max}}{w_s \times pt_{ij}.d_s})} 
 \end{equation*}
 \begin{equation*}
  = \frac{1}{\frac{1}{pss(pt_{li})} + \frac{1}{pss(pt_{ij})}}
\end{equation*}
\qed
\end{proof}

\section{Proof of Lemma~\ref{lemma:pruning}}
\label{appendix2}
\begin{proof}
It is straightforward to show that $pt^\prime_{lj}\oplus pt_{jk}$ is at least as short as $pt_{lj}\oplus pt_{jk}$ because $pt^\prime_{lj}$ is at least as short as $pt_{lj}$ and $pt_{jk}$ is same in both paths. According to Lemma~\ref{ss_property2}, the PSS of $pt^\prime_{lj}\oplus pt_{jk}$ is $\frac{1}{\frac{1}{pss(pt^\prime_{lj})} + \frac{1}{pss(pt_{ik})}}$ and the PSS of $pt_{lj}\oplus pt_{jk}$ is $\frac{1}{\frac{1}{pss(pt_{lj})} + \frac{1}{pss(pt_{ik})}}$. $\frac{1}{pss(pt_{ik})}$ part is same in the PSS of both paths. Since $pss(pt^\prime_{lj}) \geq pss(pt_{lj})$,  $\frac{1}{\frac{1}{pss(pt^\prime_{lj})} + \frac{1}{pss(pt_{ik})}} \geq \frac{1}{\frac{1}{pss(pt_{lj})} + \frac{1}{pss(pt_{ik})}}$. Thus, $pt^\prime_{lj}\oplus pt_{jk}$ is at least as safe as $pt_{lj}\oplus pt_{jk}$.
\qed
\end{proof}